\documentclass[%
reprint,
notitlepage,
superscriptaddress,
showkeys,
preprintnumbers,
nofootinbib,
amsmath,
amsfonts,
amssymb,
aps,
prd,
floatfix,
]{revtex4-2}

\usepackage{pdfpages}

\makeatletter
\AtBeginDocument{\let\LS@rot\@undefined}
\makeatother

\usepackage{float}

\DeclareGraphicsExtensions{.pdf,.png,.jpg}

\usepackage[top=1in,bottom=1in,left=1in,right=1in]{geometry}

\usepackage[mathscr]{euscript}
\usepackage{physics}

\usepackage[OT1]{fontenc}

\def\PyUL{\textsc{PyUltraLight} }
\def\PyULs{\textsc{PyUltraLight}}

\newcommand{\BH}{_\text{BH}}
\newcommand{\Sol}{_{\odot}}

\newcommand{\Tb}{\tilde{b}}
\newcommand{\LdB}{\lambda_{\text{dB}}}
\newcommand{\vecX}{\mathbf{x}}

\definecolor{ceruleanblue}{rgb}{0.16, 0.32, 0.75}

\graphicspath{{Graphics/}}

\begin{document}

\title{Dynamical Friction From Ultralight Dark Matter}
\author{Yourong Wang}
\email{yourong.f.wang@auckland.ac.nz} 
\author{Richard Easther}\email{r.easther@auckland.ac.nz} 
\affiliation{Department of Physics, The University of Auckland\\New Zealand 1010}

\date{\today} 

\begin{abstract}
We simulate the gravitational dynamics of a massive object interacting with Ultralight / Fuzzy Dark Matter (ULDM/FDM), non-relativistic quantum matter described by the Schr\"{o}dinger-Poisson equation. We first consider a point mass moving in a uniform background, and then a supermassive black hole (SMBH) moving within a ULDM soliton. After replicating simple dynamical friction scenarios to verify our numerical strategies, we demonstrate that the wake induced by a moving mass in a uniform medium may undergo gravitational collapse that dramatically increases the drag force, albeit in a scenario unlikely to be encountered astrophysically. We broadly confirm simple estimates of dynamical friction timescales for a black hole at the center of a halo but see that a large moving point mass excites coherent ``breathing modes'' in a ULDM soliton. These can lead to  ``stone skipping'' trajectories for point masses which do not  sink uniformly toward the center of the soliton, as well as  stochastic motion near the center itself. These effects will add complexity to SMBH-ULDM interactions and to SMBH mergers in a ULDM universe.

\end{abstract}

\maketitle

\section{Introduction} \label{sec:First}

We analyse a point-like massive particle  interacting with self-gravitating quantum matter. The overall investigation is motivated by the dynamics of super-massive black holes (SMBH) moving inside Ultralight Dark Matter (ULDM) halos. ULDM, also known as Fuzzy Dark Matter, is  based on non-interacting particles with de~Broglie wavelengths long enough to influence galactic dynamics on sub-kiloparsec scales \citep{Preskill1982,Abbott1982,Dine1982,Turner1983, Khlopov1985,Press1990,Hu2000, Sin_1994, Sahni2000,Matos2000,Guzman2000,Goodman_2000, Peebles_2000,Amendola_2006,Hwang2009,Marsh2016a,NIEMEYER2020}. Common realisations of this scenario are built on axions with masses in the range $10^{-20} \sim 10^{-23}$ eV.  ULDM is non-relativistic quantum matter interacting with its own Newtonian gravitational potential and  is thus governed by the nonlinear Schr\"odinger-Poisson equation.  

On large scales, ULDM resembles cold dark matter (CDM) but its quantum properties become apparent on smaller scales, modifying the expectations for intra-galactic dynamics relative to CDM~\cite{Hui2017}. Given that conventional CDM  faces a number of challenges when confronted with the small-scale properties of galaxies, these differences are the primary motivation for ULDM models, and understanding the detailed dynamics of ULDM will be key to testing these scenarios. Moreover, in addition to ULDM, self-gravitating quantum matter may arise in the very early universe \cite{Musoke:2019ima,Eggemeier:2020zeg},  hypothetical boson stars  \cite{Guzman2004,Schwabe2016,Mocz2017} and QCD axion miniclusters \cite{Eggemeier2019}, so the underlying dynamical system is relevant to wide range of astrophysical systems.

This work is complementary to that of Lancaster {\em et al.}  \cite{Lancaster2019} who give  numerical and analytic treatments of  both point-like and extended masses moving through a ULDM background. We provide more numerical detail, but focus on point masses. The uniform background case is primarily a test for our code, recovering known analytic solutions (which are analogous to a much older problem in electron propagation~\citep{LandauCoulomb}) in the limit where the self-gravity of the quantum matter is ignored and the point mass moves with constant velocity. However, a point mass moving in an otherwise undisturbed ULDM background leaves an elongated overdensity in its wake, which eventually undergoes gravitational collapse. The deep potential of the resulting overdensity then brings the moving mass to a rapid standstill.

Conversely, when a black hole interacts with a ULDM halo the central soliton has already collapsed and is supported by ``quantum pressure''. We consider a the idealized scenario of a mass in an initially circular orbit around an unperturbed soliton. We broadly confirm simple estimates of the timescales over which an orbiting mass sinks to the center of the halo. However, the moving mass excites oscillations in the  soliton independently of the dynamical friction, and the resulting motion can be complicated and stochastic. In particular, we see possible evidence that an orbiting black hole will be ``reheated'' as it interacts with the now-dynamical soliton for some parameter combinations. This appears to increase the likelihood of core-stalling in SMBH mergers in a ULDM dominated universe in a way that is distinct from the heating of black hole binaries by the granular nature of ULDM halos, described by Bar-Or {\em et al.}  \cite{Bar-Or2018}. Consequently,for both uniform backgrounds and solitonic configurations we find that 
non-perturbative  backreaction introduces qualitatively new phenomena into ULDM dynamics.

\begin{figure*}[bt]
    \centering
    \includegraphics[width = 0.75\textwidth]{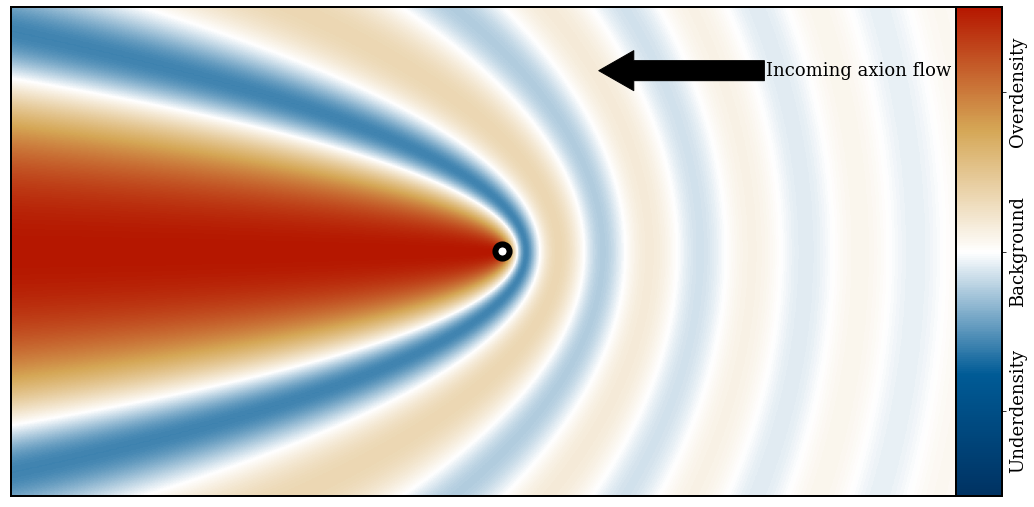}  
    \caption{The analytical density distribution for axions incident on a fixed mass (indicated by the white dot), without gravitational back-reaction. The numerical scale is omitted since the underlying equation is linear and the solution contains an undetermined multiplicative constant. }
    \label{fig:hyper}
\end{figure*}

This work rests on numerical solutions of the coupled Schr\"odinger-Poisson equation. Our simulations are based on \PyULs~\cite{Edwards2018}, a pseudo-spectral Schr\"odinger-Poisson solver written in Python (with the FFTs provided by compiled libraries) which has been modified to include point mass ensembles that react to the gravitational potential of the combined ULDM-matter system.\footnote{The code can be found at \url{https://github.com/Sifyrena/PyUL_NBody}.}

Astrophysical constraints on the axion mass  paint a complex picture \cite{Marsh:2021lqg}. Lyman-$\alpha$ forest data has been used to establish $m > 2 \times 10^{-20}$~eV at 95\% confidence \cite{Rogers:2020ltq}.  Studies of Eridanis-II \cite{Marsh:2018zyw} rule out the range $ 10^{-21}<  m < 2 \times 10^{-19}$~eV, although any given system may  be far from equilibrium  when  observed \cite{Kendall:2019fep}, complicating both ``single object'' constraints and treatments  using idealized ULDM profiles \cite{bar2021}. Conversely, superradiance \cite{Stott:2018opm} excludes masses $m \lesssim 10^{-21}$~eV but these bounds would soften in the presence of even weak self-interactions. Separately, recent large scale structure constraints \cite{Hlozek:2014lca} imply that $m>10^{-22}$~eV~\cite{Dentler:2021zij}. 
In what follows we set the axion mass to $10^{-22}$ eV for the idealised scenario of a point mass moving in a uniform ULDM background. This system has no direct astrophysical analogue and could be treated dimensionlessly. However, this value is often adopted as a fiducial ULDM mass~\citep{Ferreira2020} and providing concrete numbers contextualizes the results. When looking at interactions between point masses and ULDM solitons we set $m=10^{-21}$~eV, given that larger values are broadly preferred by the data although our overall focus here is the underlying dynamics of these systems, not their detailed astrophysics.

This paper is arranged as follows. In Section \ref{sec:ThBG}, we outline the Schr\"odinger-Poisson equation and approximate analytical treatments of dynamical friction. We describe the numerical implementation of this system in Section \ref{sec:Num}, validating the code against known results. In Section \ref{sec:Uni} we present the results for a heavy object moving in a ULDM background and the resulting gravitational collapse of the wake and we examine black hole-soliton interactions in Section \ref{sec:OneMore}.

\section{Background}\label{sec:ThBG}
\subsection{The Framework of ULDM Dynamics}

In non-relativistic limit, our system is governed by
\begin{subequations} 
\begin{align}
i\hbar\dot{\psi} &= \left[-\frac{\hbar^2}{2m}\laplacian + m(\Phi_\text{U} + \Phi_\text{N})\right]\psi,\label{eq:SoE}\\
\laplacian\Phi_\text{U} &= 4\pi Gm\abs{\psi}^2, \label{eq:PE}
\end{align}
\end{subequations}
where $\psi = \psi(\vecX,t)$ is the ULDM wavefunction, and $m$ is the axion mass. The gravitational potential due to the ULDM wavefunction is  $\Phi_\text{U}$ while $\Phi_\text{N}$ is the gravitational potential sourced by the N body particles, which themselves evolve via
\begin{subequations} 
\begin{align}
\Phi_\text{N} &= \sum_j^n \Phi_{\text{N}_j}, \label{eq:TMass}\\
\ddot{\vecX}_j &= -\sum_{k\neq j}^n \grad\Phi_{\text{N}_k}(\vecX_j) - \grad\Phi_\text{U}(\vecX_j), \label{eq:NBody}
\end{align}
\end{subequations}
Equations \ref{eq:SoE} and \ref{eq:PE} constitute the Schr\"odinger-Poisson equation with an external, time-varying potential.

Idealized ULDM halos contain a central soliton which is the ground state solution of the Schr\"odinger-Poisson equation \citep{Schive2014, Kendall2020a}. Soliton density profiles may be obtained to arbitrary numerical precision by imposing spherical symmetry on $\psi$, or
\begin{equation}
    \psi(\vecX,t) = e^{i\gamma t}f(r), \Phi(\vecX,t) = \phi(r)
\end{equation}
where $r = \abs{\vecX}$ and $\gamma$ is a constant whose value is to be numerically determined. If we define $\tilde{\phi} = \phi + \gamma$, equations \ref{eq:SoE} and \ref{eq:PE} reduce to
\begin{align}
    0 &= -\frac{1}{2} f''(r) - \frac{1}{r}f'(r) + \tilde{\phi}(r)f(r), \\
    0 &= \tilde{\phi}''(r) + \frac{2}{r}\tilde{\phi}'(r) - 4\pi f(r)^2,
\end{align}
in the time-independent limit, where $f'(r) \equiv \dv{f}{r}$ is the radial derivative. The relevant boundary conditions are $f(0) = 1$, $f'(0) = \tilde{\phi}'(0) = 0$, and $f(r_{max}) = \phi(r_{max}) = 0$ at a large enough cut-off radius $r_{max}$, which ensures that the profile is smooth at the origin.

If $e^{i\gamma t}f(r)$ is a solution to the spherically symmetric Schr\"odinger-Poisson equation, then
\begin{equation}
    e^{i \alpha \gamma t}\alpha f(\sqrt{\alpha}r),
\end{equation}
where $\alpha$ is an arbitrary scaling constant, is also a solution. It is thus straight-forward to restore physical units and initialize a 3D simulation by making appropriate choices of $\alpha$. 

\subsection{Steady State Gravitational Wakes}

As a massive object travels through a diffuse medium some of its kinetic energy and momentum may be injected into the medium. This effective drag force is known as dynamical friction. These interactions can be purely gravitational: the ``wake''  behind a moving object is over-dense and  gives rise to a force on the object opposed to its direction of motion.

Following the approach pioneered by Chandrasekhar~\citep{Chandrasekhar} it is common and usually sufficient to ignore the subsequent  evolution of the medium driven by its gravitational self-interaction. In this limit and with a constant velocity for the point mass ULDM dynamics can be viewed as Coulomb scattering \citep{LandauCoulomb,Hui2017} by working in the  frame in which a stationary mass is subject to an ``axion wind''. Consequently, we assume the particle of mass $M$ is at the origin  immersed in an axion flow with velocity $\mathbf{v} = -v_\text{rel}\hat{x}$ and density $\rho$ when undisturbed. 

Ignoring axion self-gravity and denoting the radial coordinate $\mathbf{r} = x\hat{\bf x} + y\hat{\bf y} + z\hat{\bf z}$, the system obeys the time-independent Schr\"odinger equation $E\psi =  \hat{H}\psi$,
\begin{equation}\label{eq:SoEr}
 \left[\frac{m v^2}{2} +\frac{GMm}{r} +\frac{\hbar^2}{2m}\laplacian \right]\psi(\mathbf{r}) = 0.
\end{equation}
This has an analytical solution in the form of a confluent hypergeometric function,
\begin{align}\label{eq:AnaWfn}
    \psi(\mathbf{r}) = &\sqrt{\rho} e^{\pi\beta/2 + 2\pi ix/\LdB} \abs{\Gamma(1-i\beta)} \times \nonumber \\ & \, \,  M\left[i\beta,1;i\frac{2\pi(r+x)}{\LdB}\right].
\end{align}
In Equation \ref{eq:AnaWfn}, $\LdB = {h}/({mv_\text{rel}})$ is the axion de Broglie wavelength and the inverse quantum Mach number is
\begin{equation}
     \beta = 2\pi \frac{GM}{v^2\LdB} \,  ,
\end{equation}  and we have
\begin{equation}
    M(a,b;z) = \sum_{n=0}^{\infty} \frac{a^{(n)}z^n}{b^{(n)}n!} \, ,
\end{equation} where $p^{(q)}$ is the Pochhammer symbol, 
\begin{equation}
p^{(q)} \equiv \frac{\Gamma(p+q)}{\Gamma(p)} \, .
\end{equation}
Figure \ref{fig:hyper} illustrates a typical density profile.

The dynamical friction is supplied by the gravitational field of the over-dense wake. However, a naive integral of the source over $\mathbb{R}^3$  diverges since the overdensity approaches a non-zero constant value at arbitrary large distances behind the moving mass. This problem (which stems from the unphysical assumption that the semi-infinite wake can be generated at a constant velocity within finite time) is solved by introducing  a spatial cutoff scale, $b$, the distance traveled by the mass relative to the medium. It is also helpful to expresses $b$ in units of the axion de~Broglie wavelength, denoted as $\Tb$,
\begin{equation}\label{eq:BInt}
    \Tb(t) = \frac{2\pi b}{\LdB} = \frac{m v_\text{rel}(t)}{\hbar}\int_0^{t} v_\text{rel}(t') dt',
\end{equation}

If the mass travels at constant velocity, the dynamical friction is \citep{Hui2017,Lancaster2019}  
\begin{equation}\label{eq:FAna}
    F_\text{DF} =   4\pi \bar{\rho} C(\Tb)\left(\frac{GM}{v_\text{rel}}\right)^2,
\end{equation}
where $C(\Tb)$ is a friction coefficient. The gravitational force on the mass is  $-M\partial{\Phi_U}/\partial{x}$, so approximately we have
\begin{equation}\label{eq:LanAdHoc}
    C(\Tb)  = \frac{v_\text{rel}^2}{ 4\pi \bar{\rho} G^2M} \abs{\pdv{\Phi_U}{x}(\vecX)} \, .
\end{equation}
When $\beta \ll 1$ we can extract $C(\Tb)$ from the wavefunction, Equation \ref{eq:SoEr},
\begin{equation}\label{eq:Lancaster}
    C(\Tb) = \text{Cin}(2\Tb) + \text{sinc}(2\Tb) - 1 + \mathcal{O}(\beta),
\end{equation}
where $\text{Cin}(x) \equiv \int_0^x \left[(1-\cos(t))/t\right]d t$ and $\text{sinc}(x) \equiv \sin(x)/x$. In the limit that $\Tb \ll 1$, one evaluates

\begin{equation}\label{eq:FAnaLo}
    C(\Tb) \approx \frac{1}{3}\Tb^{2}.
\end{equation}

\section{Numerical Methodology} \label{sec:Num}

\begin{figure*}[tb]
    \centering
    \includegraphics[width = 0.75\textwidth]{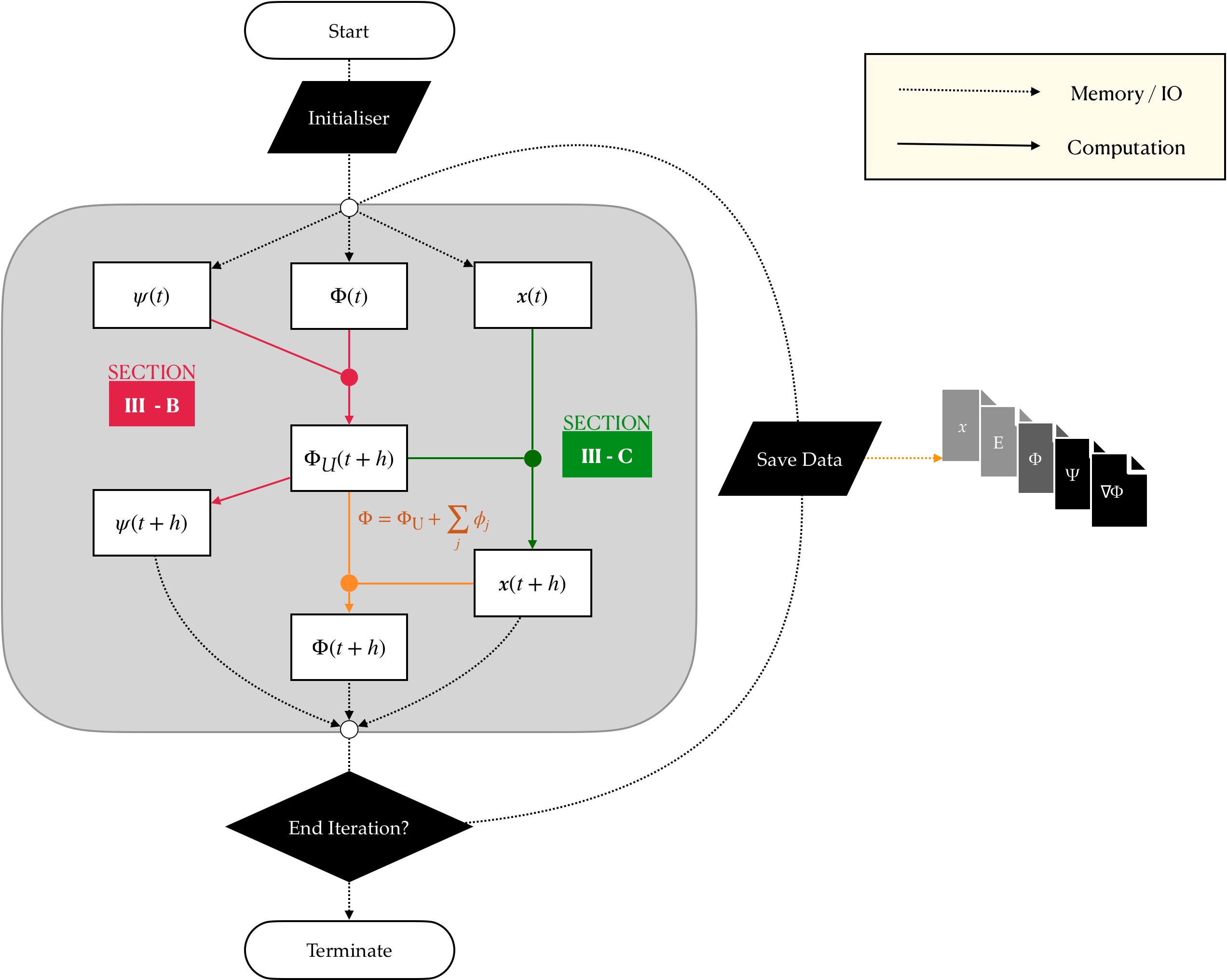}
    \caption{A flowchart of a ULDM simulation time step. The arrows' colors reflect the computational resources required: the red routine, which involves at least four 3D FFT operations, is the most expensive. $\Psi$ is the ULDM wavefunction, $\Phi$ refers to the gravitational fields, and $x$ is a vectorized representation of all particle locations and speeds.}
    \label{fig:AlgMap}
\end{figure*}

\subsection{Units and Scales} 

The program mass, time, and length units ($\mathcal{M}_c$, $\mathcal{T}_c$, and $\mathcal{L}_c$ respectively) are  as follows:
\begin{subequations}
\begin{align}
\mathcal{M}_c &=  \frac{1}{G}\sqrt[4]{\frac{3H_0^2\Omega_{m0}}{8\pi}}\left(\frac{\hbar}{m}\right)^{\frac{3}{2}} \nonumber \\ &\approx 2.227 \times 10^{7} m_{22}^{-\frac{3}{2}} M\Sol, \\
\mathcal{T}_c &= \sqrt{\frac{8\pi}{3H_0^2\Omega_{m0}}} \approx 75.1 \text{Gyr}, \\
\mathcal{L}_c &= \sqrt[4]{\frac{8\pi\hbar^{2}}{3m^2 H_0^2\Omega_{m0}}} \approx 38.36 m_{22}^{-\frac{1}{2}}
\text{ kpc},
\end{align}
\end{subequations}
where $H_0$ is the present Hubble constant, $\Omega_{m0} \approx 0.31$ is the matter fraction, and $m_{22} \equiv  {m}/{10^{-22}~ \text{eV}}$.
\subsection{ULDM Dynamics}

For a domain of edge length $L$ and resolution $N$, the simulation mesh grid involves a set of points:
\begin{equation}
    \tilde{\mathbf{x}} = -\frac{L}{2} \begin{bmatrix}
1\\
1\\
1
\end{bmatrix} + \frac{L}{N} \begin{bmatrix}
n_x\\
n_y\\
n_z
\end{bmatrix} ,
\end{equation}
where $n_x$, $n_y$, and $n_z$ are integers between $0$ and $N-1$.

To advance Equations \ref{eq:SoE} and \ref{eq:PE}, we approximate the unitary time evolution of the quantum field using the symmetrized split-step Fourier method, applied from right to left:
\begin{align}\label{eq:SS1}
   & \psi(t+h) =   \exp[-\frac{ih}{2}\Phi(t+h)] \times \nonumber \\
    & {} \, \, \mathcal{F}^{-1}
    \left\{ \exp[\frac{-ihk^2}{2}] \mathcal{F} \exp[-\frac{ih}{2}\Phi(t)]  \right\} \psi(t),
\end{align}
where $\mathcal{F}$ ($\mathcal{F}^{-1}$) denotes the (inverse) discrete Fourier transform on the grid. The ULDM gravitational potential is obtained by solving Poisson equation in the frequency domain,
\begin{equation}\label{eq:SS2}
    \Phi_\text{U}(t+h) = 4\pi \mathcal{F}^{-1}\left\{\left( -\frac{1}{k^2} \right) \mathcal{F} \left(\psi^{*}(t)\psi(t)\right)\right\}
\end{equation}
This method is correct to second order in time~\cite{Edwards:2018ccc}. 

\subsection{Time-step and Boundary Conditions}\label{sec:MrFreeman}

The ULDM velocity is manifest as the gradient in the phase of $\psi$; phase differences greater than $\pi$ radians are associated with numerical breakdown and a ``strobing'' effect that can lead to structure appearing to move in the wrong direction. By default, the integration step length, $h$, is chosen using the Courant–Friedrichs–Lewy (CFL) condition, such that an object with the highest speed resolvable by the grid travels exactly one grid interval during one time step, or
\begin{equation}\label{eq:Timestepper}
    h = \frac{L^2}{\pi N^2}.
\end{equation}
The CFL condition is a qualitative requirement in this context, given that the Schr\"odinger-Poisson equation is not a hyperbolic system \citep{Edwards2018}, but it provides a useful starting point and we have tested our results for sensitivity to the specific choice of timestep. The N body integrator takes 32 Runge-Kutta 4 (RK4) integration steps during the time $h$.

Our simulation has periodic spatial boundary conditions. For the case of a black hole moving in a uniform ULDM background we limit the duration of simulations to 
\begin{equation}
    t_\text{Max} = \frac{L}{2v_\text{rel}},
\end{equation}
so that the ULDM wake is  prevented from ``wrapping round'' the periodic boundary.  This is less of an issue when the black hole interacts with a soliton. 
 
\subsection{N Body Dynamics}

Particle potentials are implemented as Plummer spheres~\citep{Plummer1911} to suppress numerical irregularities at grid crossings\footnote{Note that this model does not capture short-range strong-field gravitational phenomena, such as superradiance.} A particle with mass $M_j$ at location $\vecX_j$ has a gravitational potential 
\begin{equation}
    {\Phi_\text{N}}_j(\vecX) = -\frac{GM_j}{\sqrt{r_P^2+r_j^2}},
\end{equation}
where $r_j = |\vecX_j-\vecX|$, and $r_P$ is the Plummer radius; for small $r_P$ this approximates an ideal point mass. 

Fourier series obtained for $\psi$ and $\Phi$ from the pseudospectral algorithm (Equations \ref{eq:SS1} and \ref{eq:SS2}) are only guaranteed to converge to the solution at the spatial grid points; evaluating them at arbitrary positions induces spurious sub-grid structure in $\Phi$. Consequently, we advance Equation \ref{eq:NBody} by estimating $\grad\Phi_U$ using a trilinear interpolation which  makes use of $\Phi$ values at the particle's $4^3$ nearest grid points. The algorithm is shown schematically in Figure~\ref{fig:AlgMap}.


\section{Dynamical Friction in a Uniform ULDM Medium} \label{sec:Uni}

\subsection{A Model Without Self-Gravity}\label{sec:NSG}

\begin{figure}[tb]
\centering
\includegraphics[width = 0.95\columnwidth]{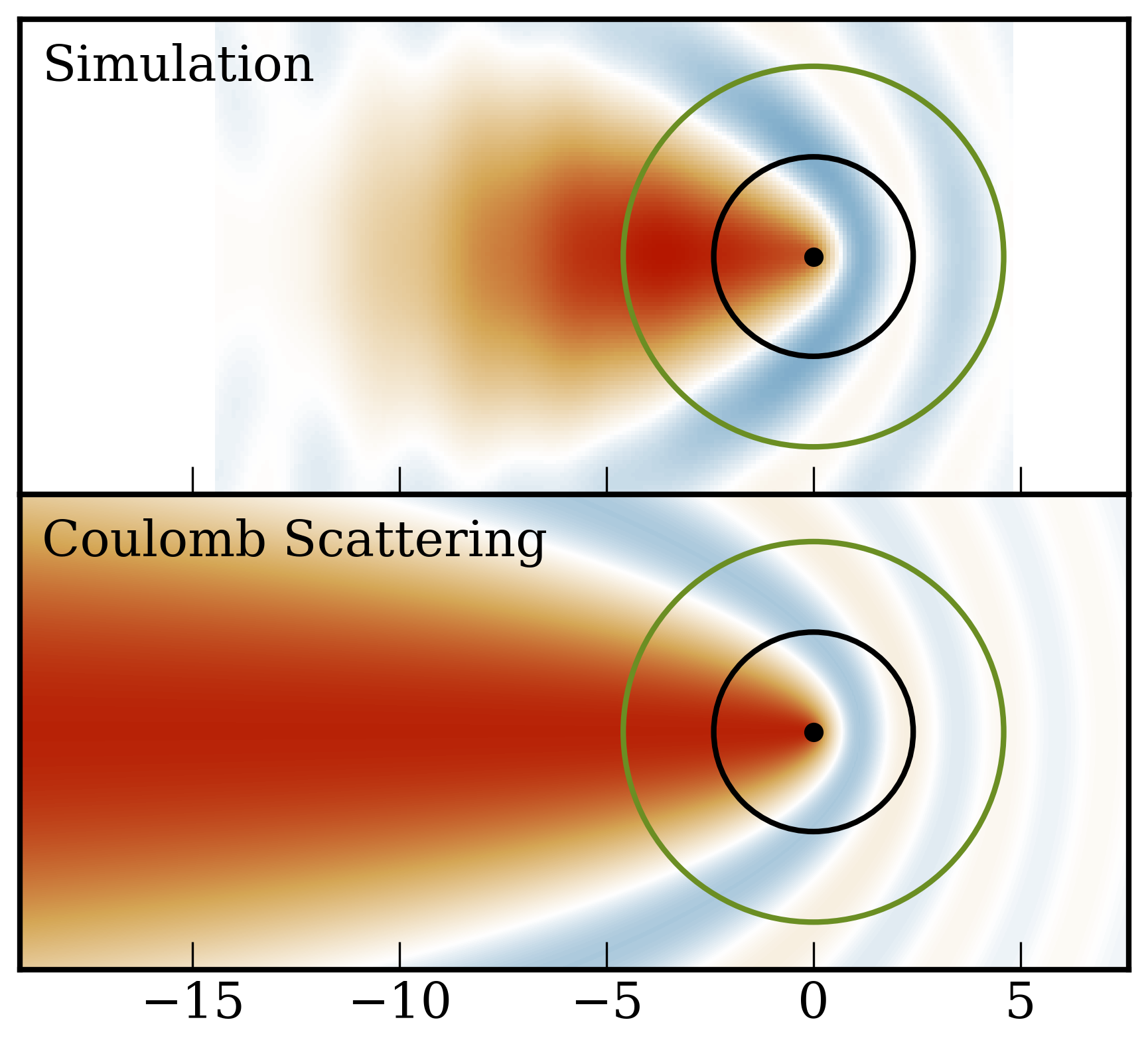}
\includegraphics[width = 0.85\columnwidth]{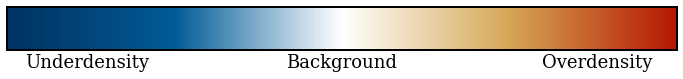}
\caption{A slice of the simulated overdensity at $192^3$ resolution without self-gravity versus the analytical Coulomb scattering result (Equation \ref{eq:AnaWfn}). The length unit are in kiloparsecs, and the circle overlays represent instantaneous values of $\LdB$ (black) and $b$ (green).}
\label{fig:HypsAtHome}
\end{figure}

We begin with simulations without ULDM self-gravity. As before, we assume a mass moving along the $x$ axis in an initially uniform ULDM medium at a constant velocity. Unless noted otherwise, the simulations shown in this section are produced with:
\begin{align*}
    L &= 4\lambda_\text{dB} \approx 9.63~ \text{kpc} \ , \\
    \rho_0 &= 10^7 \rho_\text{crit} \approx  1.27 M\Sol/\text{pc}^3\ , \\ 
    M\BH &= 10^7 M\Sol \ , \\
     v_\text{rel} &= 50~\text{km}/\text{s} \ \\
    r_P &= 48~ \text{pc}\ , \\
    \beta &= 0.0449 \ ,\\
    m_{22} &=1 \, .
\end{align*}

These quantities can be calibrated against expectations for the central solitonic condensations of ULDM halos \citep{CoreHalo}:
\begin{subequations}
\begin{align}
\rho_c = & \ 2.94\times10^{-3}M\Sol \text{pc}^{-3}\left(\frac{M_\text{vir}}{10^9M\Sol}\right)^{4/3}m_{22}^2, \\
r_c = & \ 1.6\text{kpc} \left(\frac{M_\text{vir}}{10^9 M\Sol}\right)^{-1/3}\frac{1}{m_{22}},
\end{align}
\end{subequations}
where $\rho_c$ and $r_c$ are the central density and HWHM radius of a halo with virialized mass $M_\text{vir}$. The background density in our simulations is similar to that of the solitonic core of a $10^{11} M\Sol$ halo, but our uniform-density simulated volume is substantially larger than the soliton. In Figure \ref{fig:HypsAtHome} we compare a simulation (with axion self-gravity disabled) to the  steady state Coulomb solution. There is good qualitative overlap between the two solutions in the the vicinity of the mass point. However, the wake is truncated in the numerical simulation as a consequence of the finite duration of the calculation.

\begin{figure}[tb]
    \centering
    \includegraphics[width = \columnwidth]{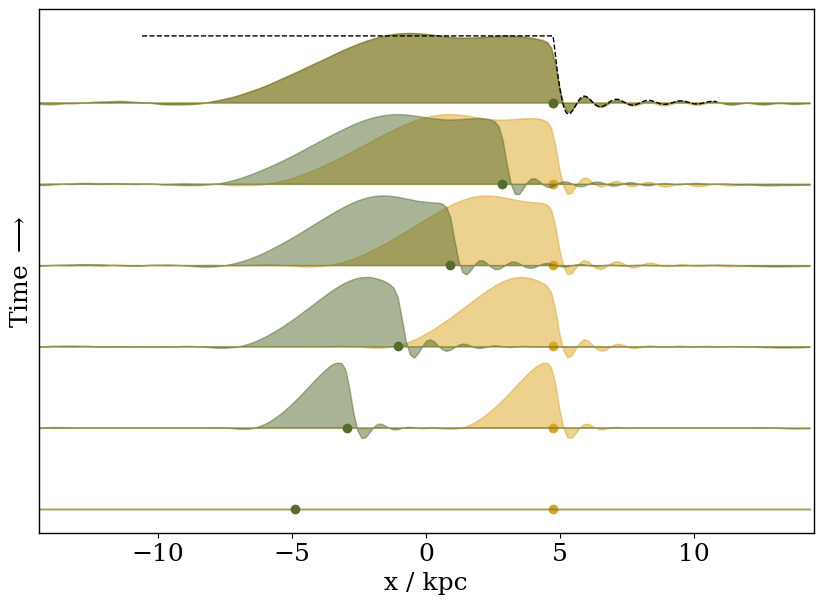}
    \caption{The overdensity along the $x$-axis in the co-moving (green) and the ULDM (yellow) frames, as described in the text. The analytical density profile due to Equation \ref{eq:AnaWfn} is superimposed on the final snapshot.}
    \label{fig:GRel}
\end{figure} 

We can work with two inertial frames, the {\it ULDM frame} and the {\it initially comoving frame}. In the former, the mass has initial velocity $\mathbf{v}_m = v_\text{rel} \hat{\mathbf{x}}$ in a stationary ULDM background. In the latter, the mass is initially at rest, embedded in a ULDM background moving with velocity  $-v_\text{rel}\hat{\mathbf{x}}$. Figure \ref{fig:GRel} illustrates that our simulations are consistent between these frames. 

\begin{figure}[tb]
\centering
\includegraphics[width = \columnwidth]{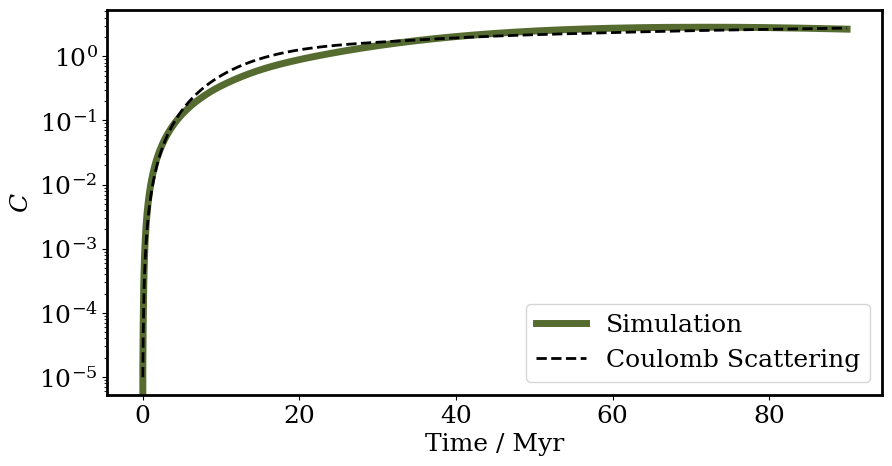}
\caption{The dynamical friction coefficient, $C$, extracted from a canonical simulation without self-gravity, plotted against time. A theoretical result obtained by substituting $b = v_\text{rel}t$ into Equation \ref{eq:Lancaster} is superimposed.
\label{fig:BCMatch}}
\end{figure}

If the dynamical friction does not alter $v_\text{rel}$ significantly, Equation \ref{eq:BInt} reduces to
\begin{equation}
    \Tb = \frac{mv^2_\text{rel}}{\hbar} t.
\end{equation}
Evaluating $C(\Tb)$ via Equation \ref{eq:Lancaster}, we can quantitatively compare the simulation with the analytical results, as shown in Figure~\ref{fig:BCMatch}. The simulation results are obtained via Equation~\ref{eq:LanAdHoc}, which is a direct measure of the force. This is a nontrivial result, in that it demonstrates that using a ``cutoff''  to compute the dynamical friction is a good match to that given by the time-dependent wake. 

\subsection{Simulations with  Self-Gravity}
\label{sec:SG}

\begin{figure}[tb]
    \centering
    \includegraphics[width = 0.9\columnwidth]{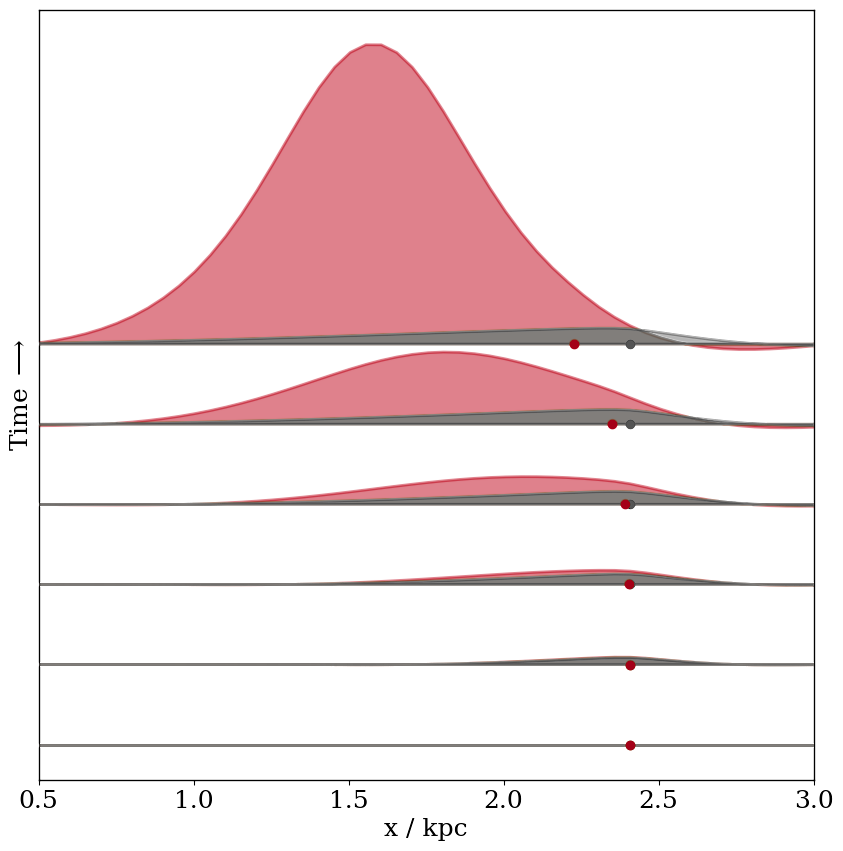}
    \caption{Time evolution of a gravitational wake behind the test mass (red), compared with   simulation result without self-gravity (dark gray). With all gravitational interactions enabled the overdense wake undergoes collapse, and the mass falls backwards (in the comoving frame) into the resulting potential.} \label{fig:DensityEvo}
\end{figure}

\begin{figure}[tb]
    \centering
    \includegraphics[width = \columnwidth]{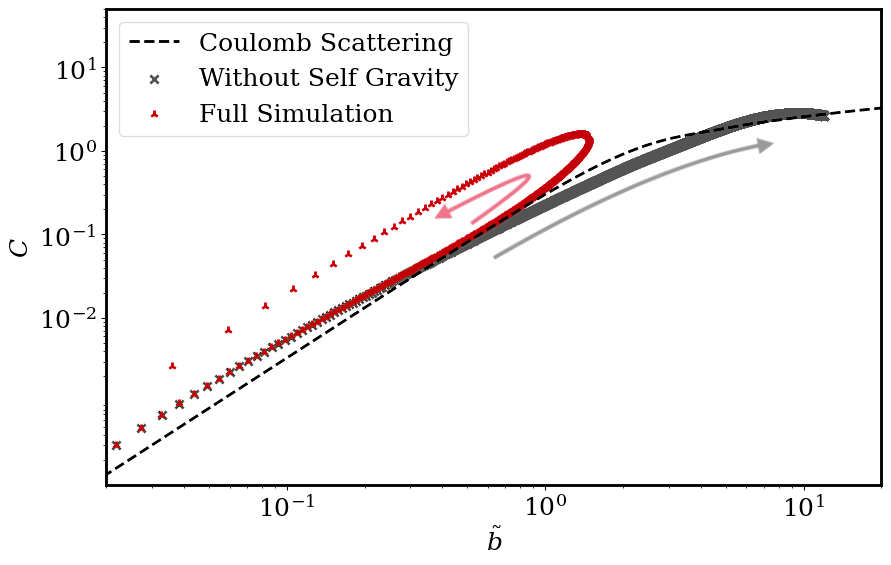}
    \caption{Dynamical friction coefficient $C$  with  gravitational backreaction for a representative case (red), compared with a simulation where backreaction is neglected (black). Initially, $\Tb$ tracks the perturbative solution and the increases as the collapse begins; the decreasing velocity reduces the de~Broglie wavelength and $\Tb$.
    The solid arrows represent the flow of time in each scenario.}
    \label{fig:LMisMatch}
\end{figure}

We now enable ULDM self-gravity and allow the traveling mass to slow down in response to the ULDM potential. In this case, the wake  undergoes gravitational collapse forming a high-density region behind the particle. The resulting gravitational potential greatly increases the dynamical friction, bringing the particle to a rapid halt.

Figure \ref{fig:DensityEvo} illustrates the time-evolution of such an overdensity. It initially tracks the previous case, but eventually tips over into a runaway collapse. We plot $C$ and $\Tb$ for our representative solution in Figure \ref{fig:LMisMatch}. Once the collapse is well underway, $v$ decreases, causing $\Tb$ to similarly decrease. 

\begin{figure}[tb]
    \centering
    \includegraphics[width=\columnwidth]{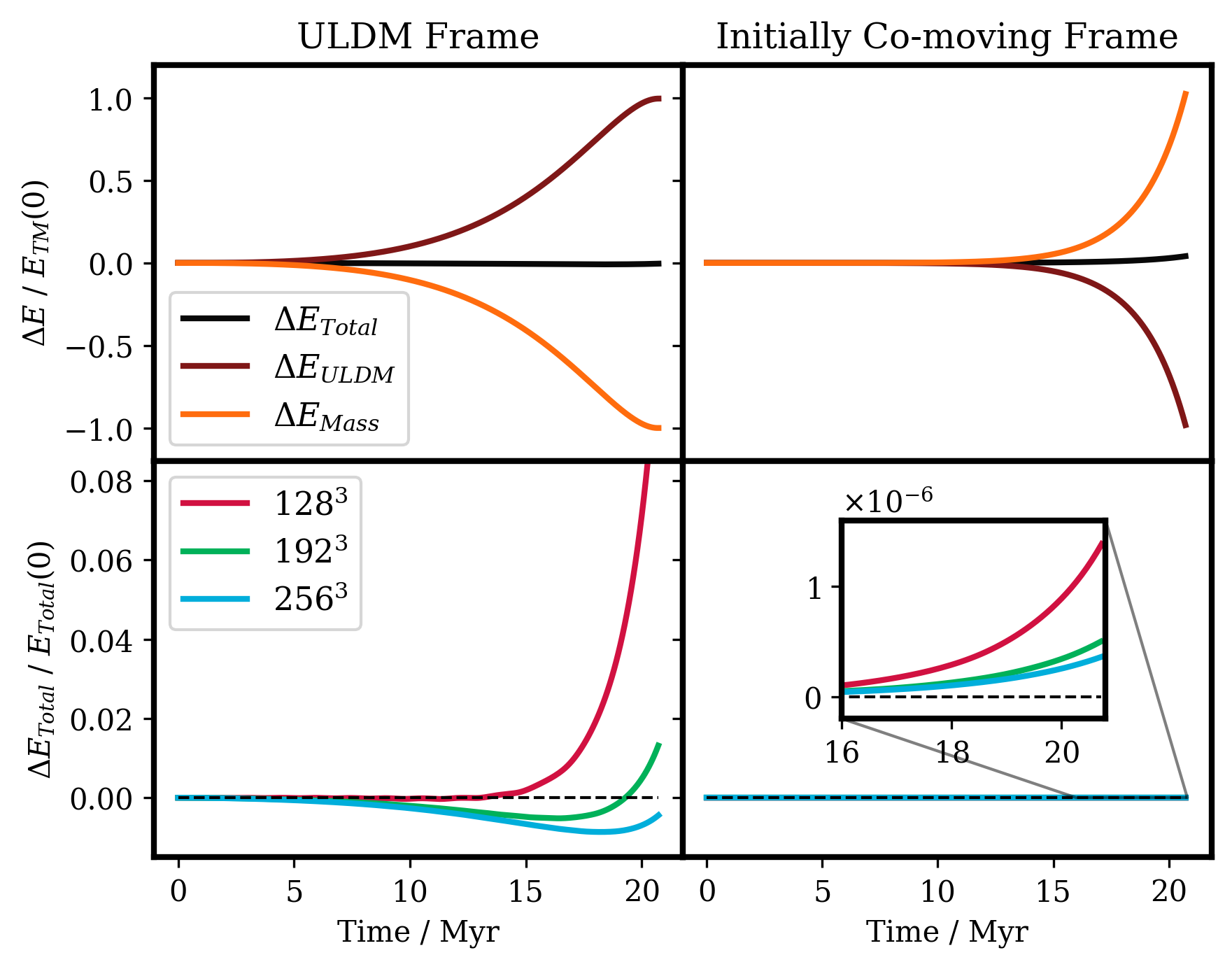}
    \caption{ \textbf{Upper Row:} The energy transfer between the travelling mass and ULDM, in units of the object's initial kinetic energy in the ULDM frame, with $N=256$. After 20 Myr the mass is sensibly at rest. \textbf{Lower Row:} Net change in system energy for $N =128$, $192$ and $256$. }
    \label{fig:EFull}
\end{figure}

In Figure \ref{fig:EFull}, we show the energy transfer between the moving mass and the background medium in the two reference frames. In the initially comoving frame the total kinetic energy is larger since a much greater mass of ULDM is moving toward the black hole, in contrast to the ULDM rest frame in which only the black hole is moving initially. In both cases we find good energy conservation, but the total amount of energy is not invariant under the Galilean transformation. In the lower plot we see that energy conservation improves with resolution as we would expect. Conservation appears to be better in comoving frame. However, this is a byproduct of the  axion flow carrying more kinetic energy than the moving mass, rather than a physical distinction. 

\begin{figure}[tb]
     \centering
    \includegraphics[width = \columnwidth]{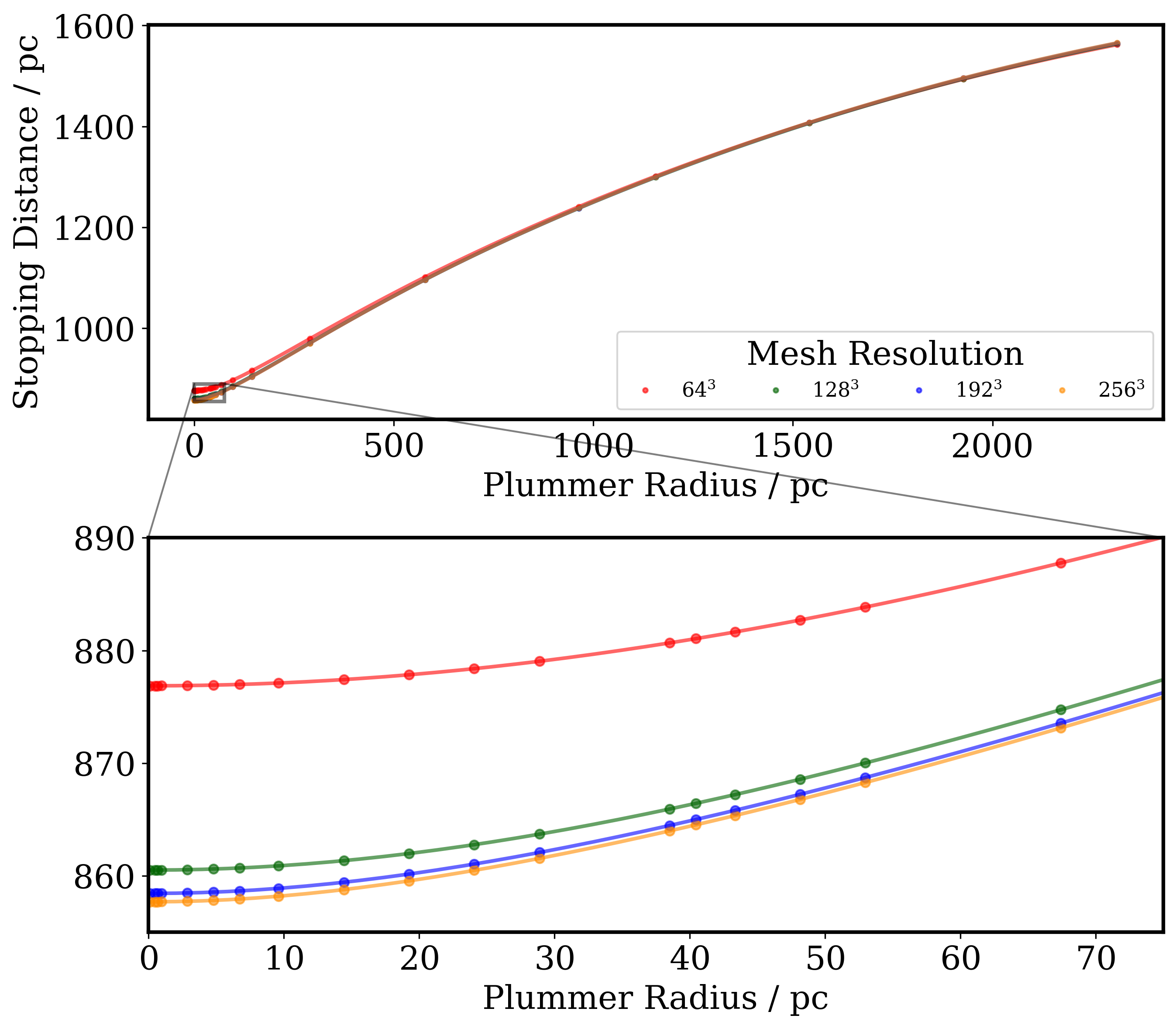}
    \caption{The stopping distances of the $10 M\Sol$ object as a function of its Plummer Radius, simulated at 4 mesh resolutions.}
    \label{fig:NaCompare}
\end{figure}

Figure \ref{fig:NaCompare} shows the dependence  on  mesh resolution and the  Plummer radius. We see that decreasing the Plummer radius increases the friction and decreases stopping distance, as expected \cite{Lancaster2019}.  We also verify  a sub-grid Plummer radius can be chosen without inducing numerical instability. Conversely, if we fix the Plummer radius relative to the grid spacing, decreasing $N$ effectively makes the potential  more diffuse, so stopping time increases as $N$ is reduced.  However, one can extrapolate to the continuum limit without difficulty. 

\begin{figure}[b]
\centering
\includegraphics[width = \columnwidth]{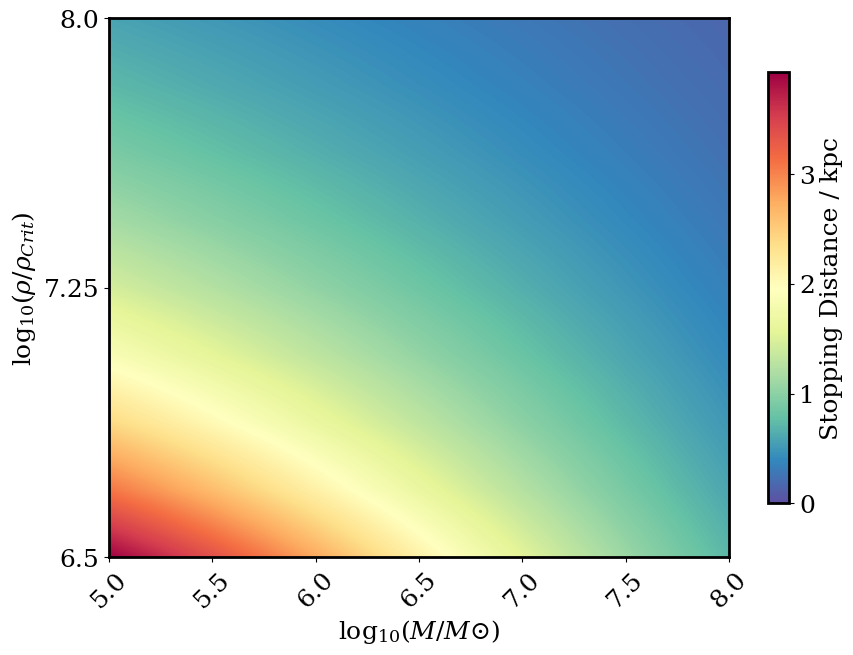}
\label{fig:SL}
\caption{Stopping distance interpolated using 13 object massed and 9 density values, all with $v_\text{Rel} = 50 $ km/s. The simulations were conducted at $128^3$ resolution in the ULDM frame.}
\end{figure}

When the self-gravity term in the Schr\"odinger-Poisson equation is small the Coulomb scattering approximation is typically sufficient to compute the force on a moving particle. However, once the wake becomes gravitationally unstable the particle  rapidly slows down.  To illustrate this we surveyed a range of initial particle masses between $0.1$ and $100$ million solar masses and ULDM densities between $10^5$ and $10^8 \rho_\text{crit}$. In almost all cases the moving mass came to halt after traveling less than 3.5 kpc and within 100 Myr.  For large black holes in a very dense ULDM background the stopping distance can be on the order of ${\cal{O}}(10)$ parsecs.

\begin{figure*}[tb]
\centering
\includegraphics[width = \textwidth]{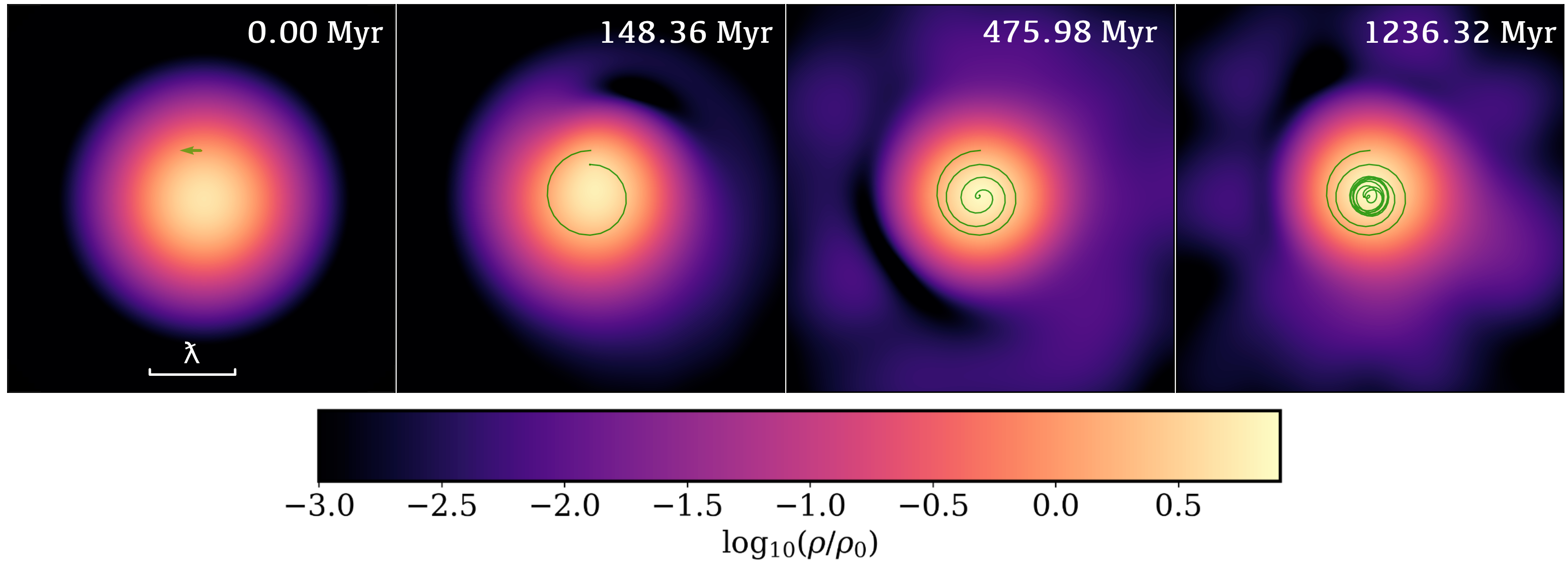}
\caption{ULDM configuration in a black hole-soliton pair with an initial separation of 300pc and $M\BH/M_\text{Soliton} = 0.08$. Density is shown on a log scale, calibrated against the value at the initial location,  $\rho_0 \approx 0.0295 M\Sol / \text{pc}^{-3}$. The de Broglie wavelength is plotted for reference.}
    \label{fig:ULDSmolLoop}
\end{figure*}

\begin{figure}[b]
    \centering
    \includegraphics[width = \columnwidth]{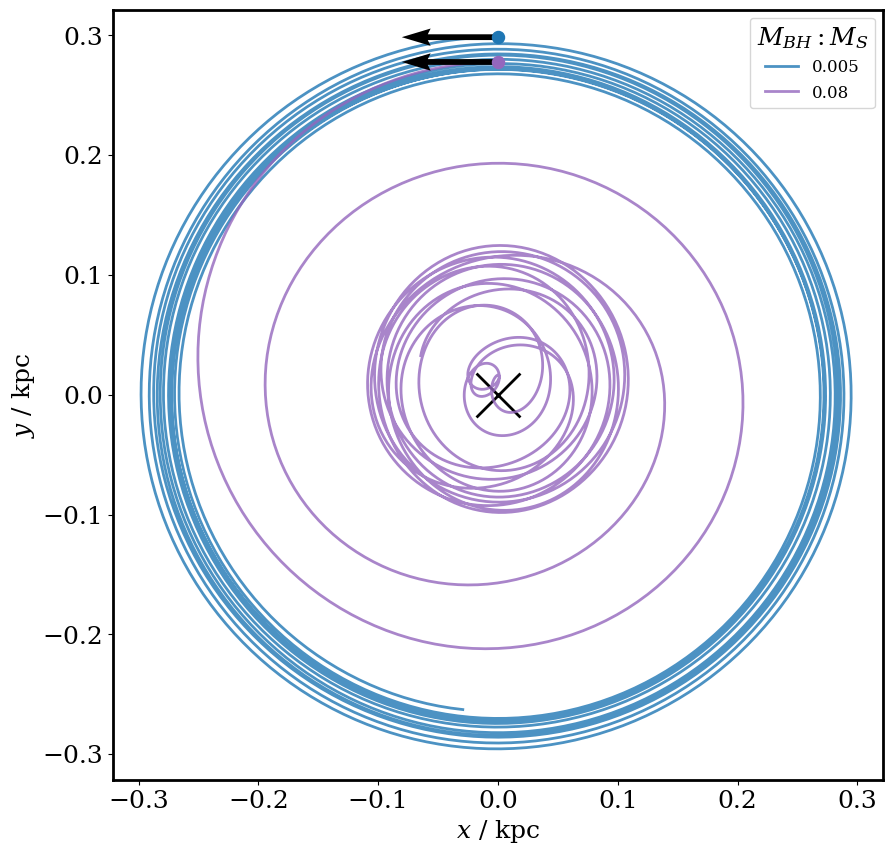} 
    \caption{Trajectories from two simulations with  initial black hole-soliton separation of 300 pc with respect to the individual system centers of mass. The orbital decay of the smaller black hole is significantly slower.}
    \label{fig:Topdown}
\end{figure}

Physically, however, this scenario is unlikely to be encountered in practice -  the densest parts of a ULDM are the central soliton, which need not behave in the same way as a uniform ULDM background. Conceivably conditions close to this scenario could exist in the early universe (recalling that $\rho \approx 10^9 \rho_\text{crit}$ at recombination) but in that scenario the moving object would necessarily be a primordial black hole, formed in a much earlier epoch. Moreover, in this scenario axion collapse may lead to the formation of a black hole, as studied in Ref.~\citep{Helfer2017}.

\section{ULDM Solitons}
\label{sec:OneMore}
\subsection{Physical Configuration}

We now consider a mass  moving in an initially circular orbit around (and inside) a Schr\"odinger-Poisson soliton and analyze the decay of its orbital radius and energy. SMBH dynamics after a galactic merger are obviously a key motivation for this work but we focus on a single, displaced SMBH in this initial treatment. 

The simulations in this Section make use of a soliton with the parameters 
\begin{align*}
    M_\text{Soliton} &= 1.2 \times 10^7 M\Sol,     \\ 
    m_{22} &= 10, \\
    r_\text{50} &= 279.7 \text{pc},
\end{align*}
where $r_\text{50}$ is the radius which encloses $50\%$ of the soliton mass.\footnote{These numbers cannot all be chosen independently; any two of them  fully specify the properties of the soliton.}   The chosen axion mass ($10^{-21}$ eV) is broadly compatible with  current astrophysical bounds (although see \cite{Rogers:2020ltq}); the mass of the central soliton is consistent with that expected for a  $\sim 10^{10} M\Sol$ halo \cite{Schive2014,Hui2017}. The simulations are performed in a box $L= 4.5$kpc on a side and the Plummer radius is set to be half of the grid-spacing.   

Our simulations begin with the black hole embedded in an undisturbed soliton. Figure~\ref{fig:ULDSmolLoop} shows the ULDM configuration at four different times for a mass ratio of $M\BH/M_\text{Soliton} = 0.08$.  There is no obvious wake, since the ULDM background responds to both quantum pressure, and its own confining gravitational potential, but the overall soliton is disturbed by the passage of the black hole.

Figure~\ref{fig:Topdown} shows the trajectories of two black holes (from separate simulations) with masses $6 \times 10^{4}$ and $9.6 \times 10^{5} M\Sol$ in initially circular orbits; the more massive black hole feels a larger dynamical friction and quickly sinks towards the center. The center of mass is at the origin, so the more massive black hole has a smaller initial radial position. 

\begin{figure}[tb]
    \centering
     \includegraphics[width = \columnwidth]{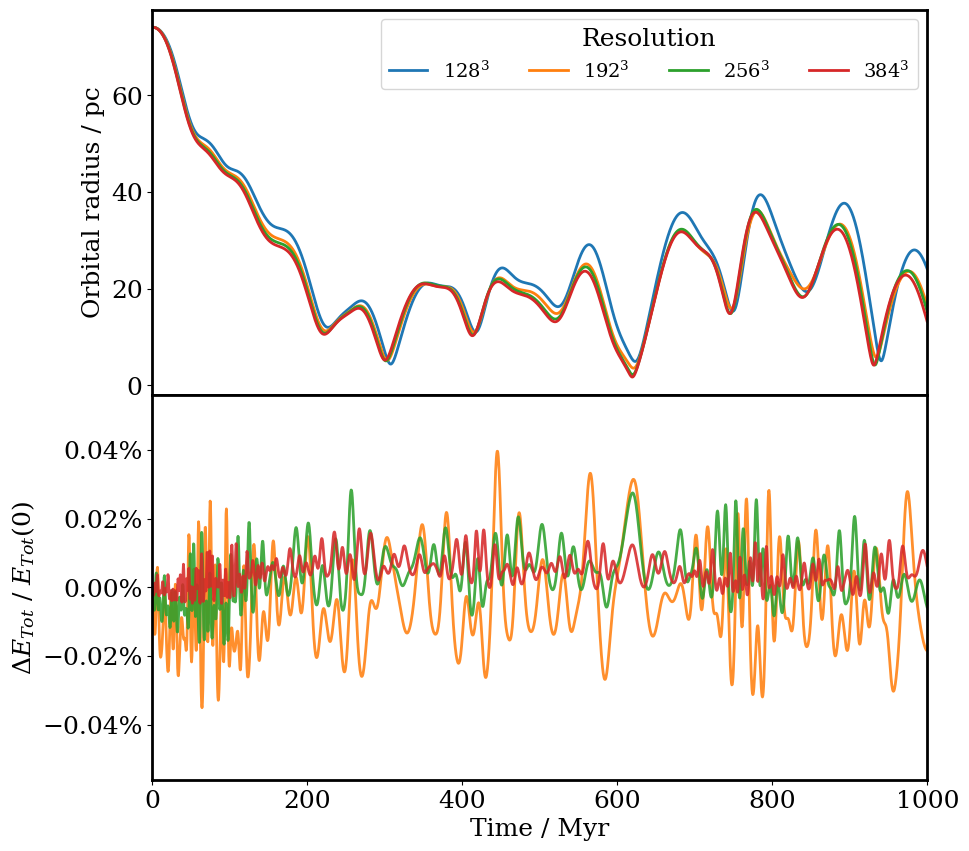}
    \caption{The black hole trajectory (top) and  energy conservation (bottom) is shown for a mass ratio of 8\% and an initial separation of 80pc for a range of resolutions.}
    \label{fig:VE}
\end{figure}

\subsection{Numerical Considerations}

These simulations are performed in \PyUL{} with periodic boundary conditions;  to suppress artifacts arising from interactions with the boundary the simulation volume is necessarily much larger than the soliton. However, our results are largely insensitive to the spatial resolution of the ULDM simulation and  energy conservation scales as expected with resolution, as shown in Figure~\ref{fig:VE}. 

The resolution-independence of these simulations is perhaps surprising, given that the whole trajectory in Figure~\ref{fig:VE} fits into a region only a few mesh grids across for $N=128$. However, this welcome result makes physical sense given that dynamical friction arises from a collective interaction between the black hole and the overall soliton, in contrast to drag forces associated with the mechanical displacement of a medium which are thus largely local phenomena. Consequently, provided the soliton is adequately resolved our simulations  quickly reach a resolution-independent limit as $N$ is increased. Recall too that the black hole position varies continuously with the lattice on which the wavefunction $\psi$ is obtained. With $N=384$, lattice points are about 11pc apart, which is on the order of the minimum radial separation attained after the black hole has sunk toward the center of the soliton.

\subsection{Dynamical Friction}

\begin{figure}[tb]
    \centering
    \includegraphics[width = 0.95\columnwidth]{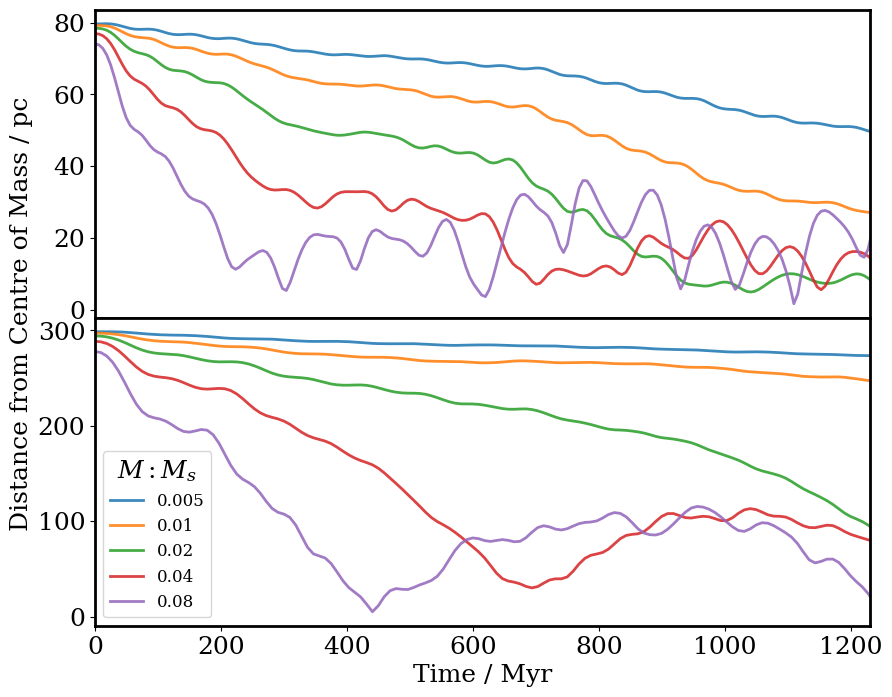} 
    \caption{Black hole orbital radii for five mass ratios. The initial radii are 80 and 300 pc in the upper and lower panels respectively; all simulations run for 1.2 billion years.}
    \label{fig:VPanels}
\end{figure}

Figure \ref{fig:VPanels} shows the trajectories for five  black hole  masses and two different starting radii.  The black holes all initially sink toward the center but their  kinetic energies need not decrease monotonically, due to their interactions with the newly disturbed soliton. 

For circular motion the dynamical friction applies a torque on the moving mass, which gives the rate of change in the angular momentum. This implicitly defines a (rough) timescale for the orbital lifetime  \citep{Hui2017}
\begin{equation}
    \tau \equiv \frac{\cal{L}}{r\abs{F_{DF}}} = \frac{1}{C} \frac{\mathfrak{M}(r)^{3/2}}{4\pi \rho M\sqrt{Gr^3}},
\end{equation}
where $\cal{L}$ is the initial orbital angular momentum and $\mathfrak{M}(r)$ is the ULDM mass inside the radius $r$. We invoke Equation \ref{eq:FAnaLo} to write
\begin{equation}\label{eq:CharaTS}
    C \approx \frac{1}{3} \Tb^2 \approx \frac{1}{3} \frac{Gm^2r\mathfrak{M}(r)}{\hbar^2}.
\end{equation}
which yields
\begin{equation}
    \tau   \approx    \frac{3 \hbar^2 \mathfrak{M}(r)^{1/2}}{4\pi m^2 \rho(r) M\sqrt{G^3r^5}}, \label{eq:timescale}
\end{equation}
where we have explicitly denoted  the density is function of $r$. Hui {\em et al.} \citep{Hui2017} assume that the black hole is near the center of the soliton and replace $\rho$ with its maximum value; after this substitution it is immediately clear that $\tau \rightarrow \infty$ as $r \rightarrow 0$.

\begin{figure}[tb]
    \centering
    \includegraphics[width = \columnwidth]{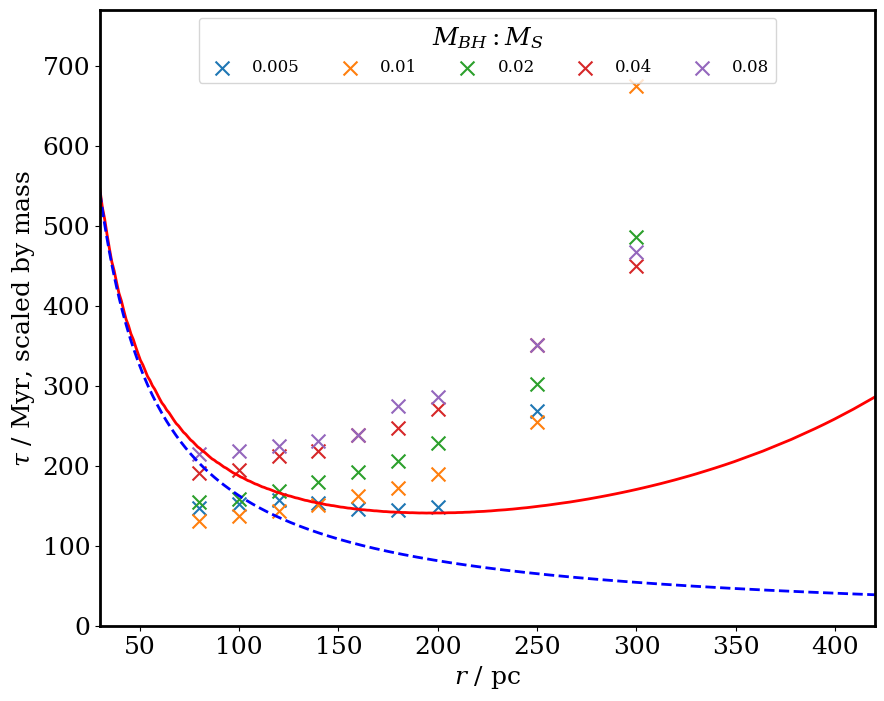}
    \caption{Orbital decay timescales for a $10^6 M\Sol$ black hole. The solid curve is based on Equation \ref{eq:timescale}; the dashed curve results from fixing the density to the central value. The data points show the timescales obtained from simulations, scaled by $10^6 M\Sol / M_{BH} $.}
    \label{fig:Circ_CTS}
\end{figure}

For a particle with mass $M_{BH}$ orbiting this specific soliton at $r_\text{50}$ the resulting timescale is
\begin{equation}\label{eq:CharaTSH}
    \tau \approx 160.18 \ \text{Myr} \left(\frac{10^6 M\Sol}{M_{BH}} \right), 
\end{equation}
recalling that $10^6 M\Sol$ is 8\% of the soliton mass, the largest ratio we consider.  Figure~\ref{fig:Circ_CTS} plots the characteristic timescale for a range of masses and radii, rescaled by $M_{BH}/{10^6}$. As noted above, $\tau$ diverges at small $r$, since the circular velocity decreases at the center of a spherical mass distribution, and likewise at large $r$ when the density of the medium and and velocity both decrease with radius, but it is roughly constant for intermediate radii. 

The derivation of the timescale in Equation \ref{eq:timescale} implicitly assumes a linear and steady decrease in angular momentum but the simulated black hole orbital radii are clearly non-monotonic. We obtain an empirical timescale for comparison purposes from the interval over which the black hole angular momentum with respect to origin decreases by 20\%\footnote{In the simulations with the smallest black holes starting from the largest radii this threshold is never actually reached; for these cases we extrapolate.}, and then rescale to the obtain the projected time to reach $L=0$. 

There is reasonable agreement between our dynamical estimates and the computed value of $\tau$, given that it is, at best, an indicative value rather than a detailed prediction. Consequently, these results can be seen as a numerical verification of the semi-analytic treatments of the dynamical friction experienced by point masses interacting with ULDM solitons, even through the classical wakes seen in the previous Section do not form in these systems.

\subsection{Soliton Backreaction}

\begin{figure}[tb]
    \centering
    \includegraphics[width = \columnwidth]{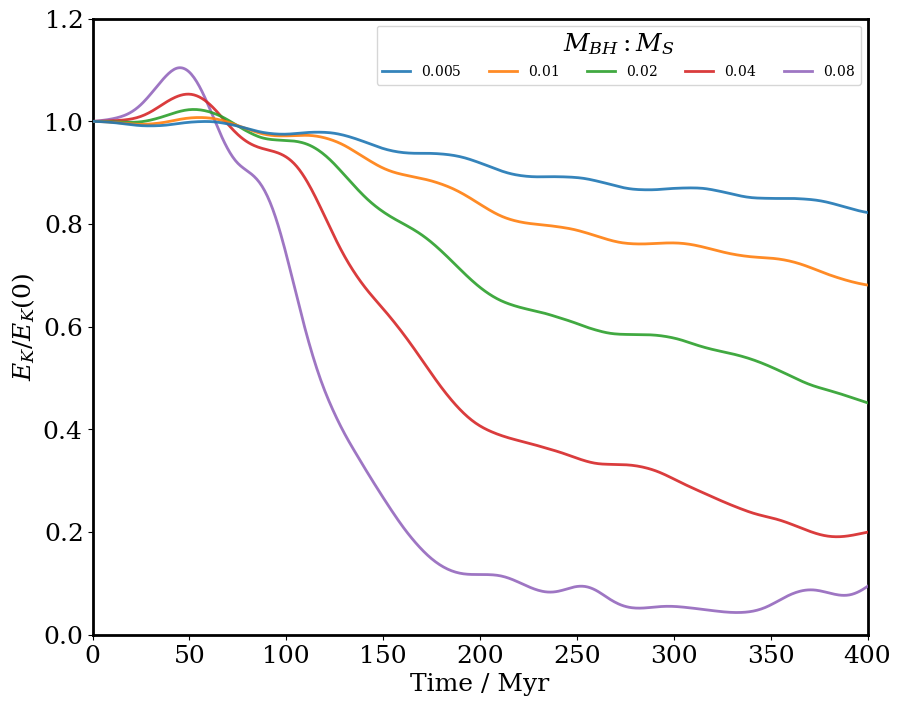}
 \caption{Kinetic energy as a function of time (relative to the initial value) for black holes with an initial radius of 80 pc.}
    \label{fig:kineticenergy}
\end{figure}

In principle, the approximation in Equation~\ref{eq:timescale} could be improved by integrating the instantaneous torque to yield the time taken to move between any two given radii. This would be less valuable in practice, given that in many cases the orbits are far from circular. For our chosen configuration, the black hole faces a force opposed to its initial velocity causing it to ``fall'' toward the center,  accelerating as it does so, as illustrated in Figure~\ref{fig:kineticenergy}. More massive black holes follow a clearly spiral trajectory toward the center, as seen in Figure~\ref{fig:Topdown}, and can undergo effectively stochastic motion upon their arrival in the central region of the soliton. This motion is reminicent of the  ``reheating'' experienced by a massive particle when it is introduced to the centers of an already excited soliton \citep{YaleDDC}.

\begin{figure}[tb]
    \centering
    \includegraphics[width = \columnwidth]{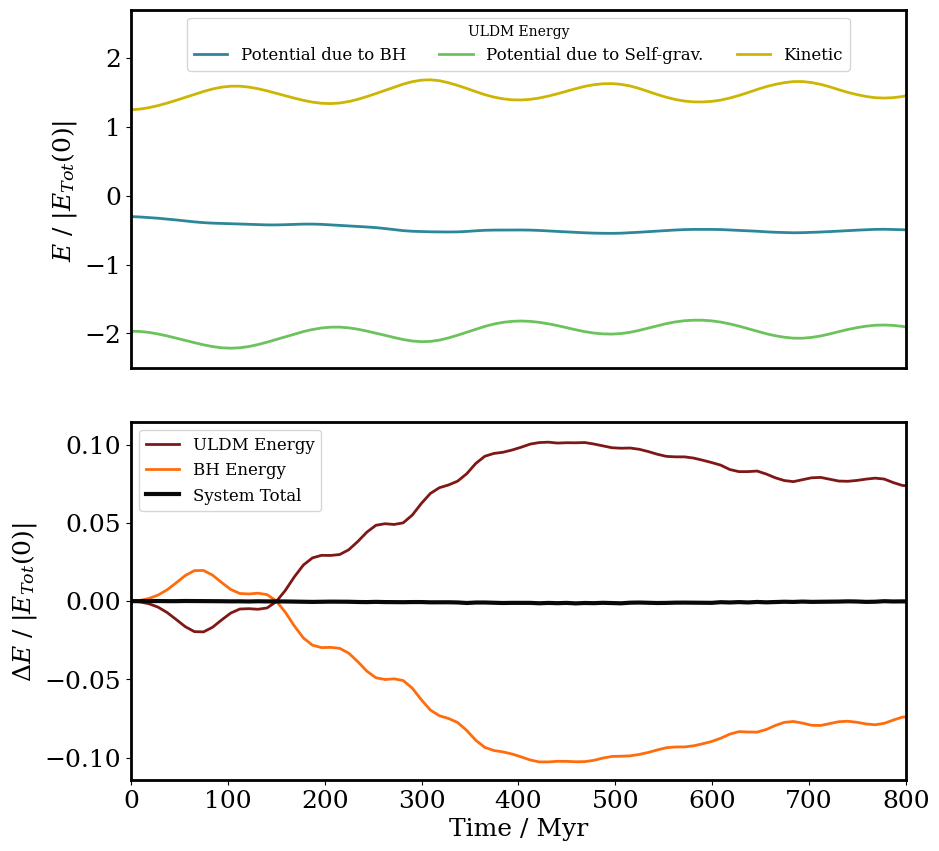}
    \caption{Components of the total energy for a simulation with $r_0 = 300$ pc and $M_{BH}/M_{Sol} = 0.08$. The top panel shows a component-wise breakdown of the ULDM energies, relative to the system's total energy. The bottom panel shows the changes in ULDM, BH, and total energies.}
    \label{fig:SolitonEnergy}
\end{figure}

\begin{figure}[tb]
    \centering
    \includegraphics[width = \columnwidth]{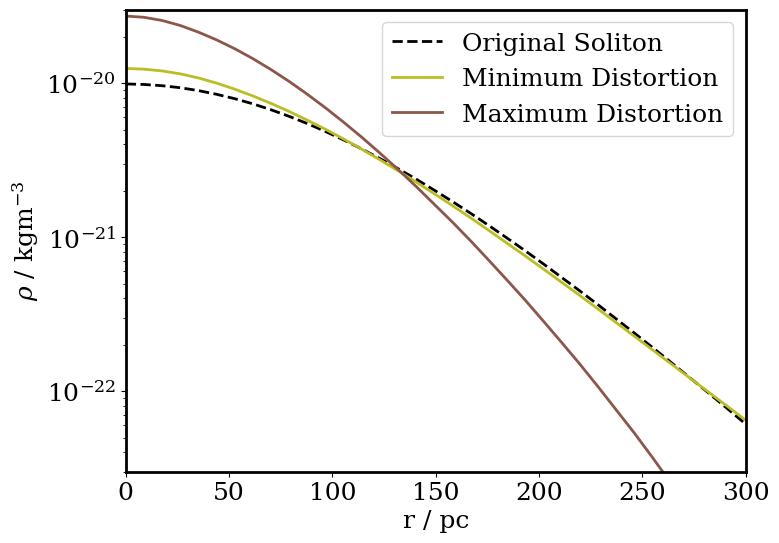}
    \caption{An idealized representation of the ULDM soliton's breathing behavior, as excited by our massive particle of $8\%M_S$, initially orbiting 80 parsecs away from the soliton's center. The soliton's density profile oscillates between the two solid lines on this graph, while the dashed line corresponds to the unperturbed soliton profile. The solid-line profiles were obtained from the simulations, via radial averaging around the ULDM center of mass.}
    \label{fig:SolBreathing}
\end{figure}

\begin{figure}[tb]
    \centering
    \includegraphics[width = \columnwidth]{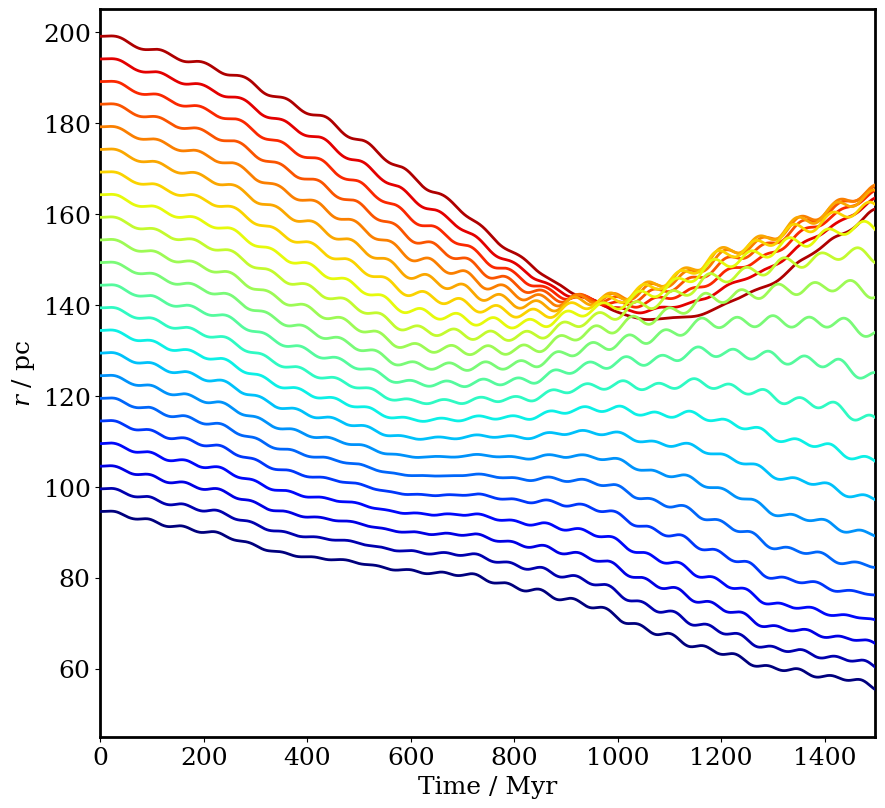} 
    \caption{Radius as a function of time for a  $0.5\% M_S$ black hole.  A trajectory with an initial separation of 200 pc actually sinks  faster than one that begins at 150 pc.}
    \label{fig:SkipStone}
\end{figure}

The individual components of the total energy for a simulation with $r_0 = 300$ pc and $M_{BH}/M_{Sol} = 0.08$ are shown in Figure~\ref{fig:SolitonEnergy}.  The overall energy of the black hole decreases as it sinks towards the center of the soliton. However, we also see the onset of a persistent  oscillation in the soliton itself, even though its total energy is constant, outside of the energy injected by the moving black hole. This is attributable to our chosen initial configuration which puts a stationary, spherically symmetric soliton in the potential of an adjacent black hole. This is a small perturbation to the overall gravitational potential of the soliton, but it means that it is no longer in its ground state configuration. The soliton is also relatively ``compressible'' -- the overall change in its self-potential is several times larger than the potential energy of the black hole. T
he impact of the breathing mode  on the potential is illustrated in Figure~\ref{fig:SkipStone}, which shows the trajectories of black holes for a series of different starting radii.  

Physically, this is a breathing mode, albeit one likely to break spherical symmetry given the off-center position of the external gravitational field. The oscillations persist on timescales much longer than those over which the black hole orbit decays as there is no mechanism to remove this energy from the system. Moreover, they persist even if the black hole is deleted from the simulation after it has completed a number of orbits.

Beyond the stochastic motion seen at the center of the soliton, for certain parameter choices the breathing mode ``reheats'' black holes orbiting at some distance from the center.  This is illustrated in Figure~\ref{fig:SkipStone}, which shows a set of trajectories in which the radial distance of the black hole  steadily increases over a number of orbits. Physically, this behavior appears to be driven by a resonance between the soliton breathing mode and the orbital period; a similar situation is described in Ref.~\citep{LisbonA}. It is more pronounced for small black holes (since it has to work against the dynamical friction, which increases with mass) and depends non-linearly on the initial radius, which fixes the specific form of the breathing mode.  This behavior (which is reminiscent of a stone skipped across a pond) is responsible for much of the scatter seen in Figure~\ref{fig:Circ_CTS}.

\section{Conclusion and Discussions} \label{sec:Disc}

We have presented simulations of (large) point masses interacting with ultralight dark matter (ULDM), and focused on two scenarios -- a uniform background of ULDM and the soliton found at the center of a ULDM galactic halo. In the former, the wake left by the moving point mass can collapse under its self-gravity, dramatically enhancing the dynamical friction. We then simulate the dynamics of a super-massive black hole in an initially circular orbit about (and inside of) a ULDM soliton. The black hole sinks towards the center of the soliton. We confirm simple estimates of the relevant timescale within ${\cal{O}}(1)$ but also see novel ``stone skipping'' trajectories at certain large initial radii, where the black hole does not monotonically approach the center. Once near the center, black holes undergo stochastic motion, migrating back out to radii of 10s of parsecs in the examples we study. Both the stone skipping and the stochastic trajectories are driven by excitations to the soliton sourced by its interactions with black hole.  

Astrophysically, there may be few circumstances in which a point mass will encounter a uniform and otherwise unperturbed background, although one can  imagine possible scenarios involving primordial black holes or very early universe physics \cite{Musoke:2019ima,Eggemeier:2020zeg}. Conversely, a massive object inside a Schr\"{o}dinger-Poisson soliton maps directly to the dynamics of SMBH at the center of a galactic halo, and these systems have a wide range of astrophysical consequences. Identifying the ways in which the distinctive properties of ULDM modify our expectations for these interactions could be key to testing the scenario, given the potential of  pulsar timing experiments \cite{Burke-Spolaor:2018bvk} and the upcoming LISA mission~\cite{LISA:2017pwj}. 

This investigation focused on a single black hole interacting with an initially unperturbed soliton but the interactions between two (or more) SMBH in a single, post-merger halo are of particular interest. These systems are fascinating in their own right, given that  most large galaxies at low-redshifts have a single central SMBH but are likely to be the products of mergers in the evolving universe. Consequently, it appears that the merger dynamics of SMBH must, to some extent, recapitulate the merger trees of their host galaxies. However, the actual processes that bring SMBH close enough to ensure that gravitational wave emission drives mergers on timescales less than the present age of the Universe are poorly understood.  Consequently, determining whether ULDM can solve  -- or exacerbate -- this so-called ``final parsec problem''  \cite{Milosavljevic:2002bn,Barausse:2020gbp} is a promising strategy for testing the overall scenario. Applying the tools developed here to the dynamics of multiple SMBH interacting within ULDM solitons is an obvious extension of this work. 

The present results complement suggestions by Bar-Or {\em et al.\/}  \cite{Bar-Or2018} that black hole binaries will be ``heated'' by  interactions with a granular ULDM halo. In the present case the ULDM is initially uniform and large scale oscillations are induced as the soliton-SMBH system orbits its common center of mass. In a post-merger halo, the central soliton may be far from its ground state, suggesting that these effects might be substantially enhanced in  astrophysical settings, resulting in the outward diffusion of light objects residing in the center of the soliton \citep{YaleDDC}. In addition, the coupling and impulsive heating associated with a single SMBH-soliton interaction could be analyzed in detail using eigenstate expansions of the soliton potential \cite{Zagorac:2021qxq}, facilitating the semi-analytic treatment of these systems. 

Perhaps surprisingly it seems that the interactions between ULDM solitons and the black hole motion are well-modeled even at low grid resolutions.  This rather fortunate outcome arises from the difficulty of establishing large density gradients in ULDM on scales significantly shorter than the de Broglie wavelength; the black hole effectively interacts with the overall soliton, rather than just the matter in its immediate locality. That said, there is clear value in high-resolution simulations. However,  we are obliged to simulate a large volume to prevent the soliton from being disrupted by boundary effects so the black hole trajectory is confined to a small fraction of the total simulation region. Consequently, implementing the combination of a hard $N$-body solver coupled to a Schr\"{o}dinger-Poisson solver in a scheme that supports adaptive mesh refinement (e.g. Ref.~\cite{Schwabe2020}) is a logical next step. 

We see interesting interactions at larger radii driven by ``breathing modes'' of the soliton excited by its interaction with the black hole. In these cases black holes at relatively large distances do not sink monotonically toward the center of the soliton. These ``stone skipping'' trajectories differ from previous work on the dynamical friction in ULDM (e.g. \cite{Lancaster2019,Bar-Or2018} in that they represent interactions between the point mass and the overall soliton, and point to further novel behaviors associated with SMBH-ULDM dynamics. 

The breathing modes driving the stone skipping solutions are reminiscent of \textit{quasinormal modes} arising from displacements of a Schr\"odinger-Poisson system away from its equilibrium  configuration \citep{Guzman2004, Zagorac:2021qxq}. These analyses can presumably be generalized to the asymmetric states seen here, allowing a more  quantitative understanding of these trajectories.

In summary, this work explores the dynamical friction acting on a massive point particle traversing through self-gravitating quantum matter, and investigates interactions between black holes and Schr\"{o}dinger-Poisson solitons. This  creates a pathway toward the detailed study of these systems in contexts ranging from boson stars, ultralight dark matter, to the early universe.

\begin{acknowledgments}
 We thank Katy Clough, Mateja Gosenca, Lilian Guo, Peter Hayman, Shaun Hotchkiss, Emily Kendall, Priyamvada Natarajan, Jens Niemeyer,  Nikhil Padmanabhan, and Luna Zagorac for useful discussions.  
  We acknowledge support from the Marsden Fund of the Royal Society of New Zealand.  The simulations in this paper were performed on the Australian National eResearch Collaboration Tools and Resources (NeCTAR) platform, hosted at the University of Auckland. The authors  acknowledge the use of New Zealand eScience Infrastructure
(NeSI) high performance computing facilities and consulting support as part of this research.
\end{acknowledgments}

\bibliographystyle{apsrev4-1}
\bibliography{main}

\begin{thebibliography}{52}%
\makeatletter
\providecommand \@ifxundefined [1]{%
 \@ifx{#1\undefined}
}%
\providecommand \@ifnum [1]{%
 \ifnum #1\expandafter \@firstoftwo
 \else \expandafter \@secondoftwo
 \fi
}%
\providecommand \@ifx [1]{%
 \ifx #1\expandafter \@firstoftwo
 \else \expandafter \@secondoftwo
 \fi
}%
\providecommand \natexlab [1]{#1}%
\providecommand \enquote  [1]{``#1''}%
\providecommand \bibnamefont  [1]{#1}%
\providecommand \bibfnamefont [1]{#1}%
\providecommand \citenamefont [1]{#1}%
\providecommand \href@noop [0]{\@secondoftwo}%
\providecommand \href [0]{\begingroup \@sanitize@url \@href}%
\providecommand \@href[1]{\@@startlink{#1}\@@href}%
\providecommand \@@href[1]{\endgroup#1\@@endlink}%
\providecommand \@sanitize@url [0]{\catcode `\\12\catcode `\$12\catcode
  `\&12\catcode `\#12\catcode `\^12\catcode `\_12\catcode `\%12\relax}%
\providecommand \@@startlink[1]{}%
\providecommand \@@endlink[0]{}%
\providecommand \url  [0]{\begingroup\@sanitize@url \@url }%
\providecommand \@url [1]{\endgroup\@href {#1}{\urlprefix }}%
\providecommand \urlprefix  [0]{URL }%
\providecommand \Eprint [0]{\href }%
\providecommand \doibase [0]{http://dx.doi.org/}%
\providecommand \selectlanguage [0]{\@gobble}%
\providecommand \bibinfo  [0]{\@secondoftwo}%
\providecommand \bibfield  [0]{\@secondoftwo}%
\providecommand \translation [1]{[#1]}%
\providecommand \BibitemOpen [0]{}%
\providecommand \bibitemStop [0]{}%
\providecommand \bibitemNoStop [0]{.\EOS\space}%
\providecommand \EOS [0]{\spacefactor3000\relax}%
\providecommand \BibitemShut  [1]{\csname bibitem#1\endcsname}%
\let\auto@bib@innerbib\@empty
\bibitem [{\citenamefont {Preskill}\ \emph {et~al.}(1983)\citenamefont
  {Preskill}, \citenamefont {Wise},\ and\ \citenamefont
  {Wilczek}}]{Preskill1982}%
  \BibitemOpen
  \bibfield  {author} {\bibinfo {author} {\bibfnamefont {J.}~\bibnamefont
  {Preskill}}, \bibinfo {author} {\bibfnamefont {M.~B.}\ \bibnamefont {Wise}},
  \ and\ \bibinfo {author} {\bibfnamefont {F.}~\bibnamefont {Wilczek}},\ }\href
  {\doibase 10.1016/0370-2693(83)90637-8} {\bibfield  {journal} {\bibinfo
  {journal} {Phys. Lett. B}\ }\textbf {\bibinfo {volume} {120}},\ \bibinfo
  {pages} {127} (\bibinfo {year} {1983})}\BibitemShut {NoStop}%
\bibitem [{\citenamefont {Abbott}\ and\ \citenamefont
  {Sikivie}(1983)}]{Abbott1982}%
  \BibitemOpen
  \bibfield  {author} {\bibinfo {author} {\bibfnamefont {L.}~\bibnamefont
  {Abbott}}\ and\ \bibinfo {author} {\bibfnamefont {P.}~\bibnamefont
  {Sikivie}},\ }\href {\doibase 10.1016/0370-2693(83)90638-X} {\bibfield
  {journal} {\bibinfo  {journal} {Phys. Lett. B}\ }\textbf {\bibinfo {volume}
  {120}},\ \bibinfo {pages} {133} (\bibinfo {year} {1983})}\BibitemShut
  {NoStop}%
\bibitem [{\citenamefont {Dine}\ and\ \citenamefont
  {Fischler}(1983)}]{Dine1982}%
  \BibitemOpen
  \bibfield  {author} {\bibinfo {author} {\bibfnamefont {M.}~\bibnamefont
  {Dine}}\ and\ \bibinfo {author} {\bibfnamefont {W.}~\bibnamefont
  {Fischler}},\ }\href {\doibase 10.1016/0370-2693(83)90639-1} {\bibfield
  {journal} {\bibinfo  {journal} {Phys. Lett. B}\ }\textbf {\bibinfo {volume}
  {120}},\ \bibinfo {pages} {137} (\bibinfo {year} {1983})}\BibitemShut
  {NoStop}%
\bibitem [{\citenamefont {Turner}(1983)}]{Turner1983}%
  \BibitemOpen
  \bibfield  {author} {\bibinfo {author} {\bibfnamefont {M.~S.}\ \bibnamefont
  {Turner}},\ }\href {\doibase 10.1103/PhysRevD.28.1243} {\bibfield  {journal}
  {\bibinfo  {journal} {Phys. Rev. D}\ }\textbf {\bibinfo {volume} {28}},\
  \bibinfo {pages} {1243} (\bibinfo {year} {1983})}\BibitemShut {NoStop}%
\bibitem [{\citenamefont {{Khlopov}}\ \emph {et~al.}(1985)\citenamefont
  {{Khlopov}}, \citenamefont {{Malomed}},\ and\ \citenamefont
  {{Zeldovich}}}]{Khlopov1985}%
  \BibitemOpen
  \bibfield  {author} {\bibinfo {author} {\bibfnamefont {M.~I.}\ \bibnamefont
  {{Khlopov}}}, \bibinfo {author} {\bibfnamefont {B.~A.}\ \bibnamefont
  {{Malomed}}}, \ and\ \bibinfo {author} {\bibfnamefont {I.~B.}\ \bibnamefont
  {{Zeldovich}}},\ }\href {\doibase 10.1093/mnras/215.4.575} {\bibfield
  {journal} {\bibinfo  {journal} {Mon. Not. R. Astron. Soc.}\ }\textbf
  {\bibinfo {volume} {215}},\ \bibinfo {pages} {575} (\bibinfo {year}
  {1985})}\BibitemShut {NoStop}%
\bibitem [{\citenamefont {Press}\ \emph {et~al.}(1990)\citenamefont {Press},
  \citenamefont {Ryden},\ and\ \citenamefont {Spergel}}]{Press1990}%
  \BibitemOpen
  \bibfield  {author} {\bibinfo {author} {\bibfnamefont {W.~H.}\ \bibnamefont
  {Press}}, \bibinfo {author} {\bibfnamefont {B.~S.}\ \bibnamefont {Ryden}}, \
  and\ \bibinfo {author} {\bibfnamefont {D.~N.}\ \bibnamefont {Spergel}},\
  }\href {\doibase 10.1103/PhysRevLett.64.1084} {\bibfield  {journal} {\bibinfo
   {journal} {Phys. Rev. Lett.}\ }\textbf {\bibinfo {volume} {64}},\ \bibinfo
  {pages} {1084} (\bibinfo {year} {1990})}\BibitemShut {NoStop}%
\bibitem [{\citenamefont {Hu}\ \emph {et~al.}(2000)\citenamefont {Hu},
  \citenamefont {Barkana},\ and\ \citenamefont {Gruzinov}}]{Hu2000}%
  \BibitemOpen
  \bibfield  {author} {\bibinfo {author} {\bibfnamefont {W.}~\bibnamefont
  {Hu}}, \bibinfo {author} {\bibfnamefont {R.}~\bibnamefont {Barkana}}, \ and\
  \bibinfo {author} {\bibfnamefont {A.}~\bibnamefont {Gruzinov}},\ }\href
  {\doibase 10.1103/physrevlett.85.1158} {\bibfield  {journal} {\bibinfo
  {journal} {Phys. Rev. Lett.s}\ }\textbf {\bibinfo {volume} {85}},\ \bibinfo
  {pages} {1158} (\bibinfo {year} {2000})}\BibitemShut {NoStop}%
\bibitem [{\citenamefont {Sin}(1994)}]{Sin_1994}%
  \BibitemOpen
  \bibfield  {author} {\bibinfo {author} {\bibfnamefont {S.-J.}\ \bibnamefont
  {Sin}},\ }\href {\doibase 10.1103/physrevd.50.3650} {\bibfield  {journal}
  {\bibinfo  {journal} {Phys. Rev. D}\ }\textbf {\bibinfo {volume} {50}},\
  \bibinfo {pages} {3650} (\bibinfo {year} {1994})}\BibitemShut {NoStop}%
\bibitem [{\citenamefont {{Sahni}}\ and\ \citenamefont
  {{Wang}}(2000)}]{Sahni2000}%
  \BibitemOpen
  \bibfield  {author} {\bibinfo {author} {\bibfnamefont {V.}~\bibnamefont
  {{Sahni}}}\ and\ \bibinfo {author} {\bibfnamefont {L.}~\bibnamefont
  {{Wang}}},\ }\href {\doibase 10.1103/PhysRevD.62.103517} {\bibfield
  {journal} {\bibinfo  {journal} {\prd}\ }\textbf {\bibinfo {volume} {62}},\
  \bibinfo {eid} {103517} (\bibinfo {year} {2000})},\ \Eprint
  {http://arxiv.org/abs/astro-ph/9910097} {arXiv:astro-ph/9910097 [astro-ph]}
  \BibitemShut {NoStop}%
\bibitem [{\citenamefont {{Matos}}\ \emph {et~al.}(2000)\citenamefont
  {{Matos}}, \citenamefont {{Guzm{\'a}n}},\ and\ \citenamefont
  {{Ure{\~n}a-L{\'o}pez}}}]{Matos2000}%
  \BibitemOpen
  \bibfield  {author} {\bibinfo {author} {\bibfnamefont {T.}~\bibnamefont
  {{Matos}}}, \bibinfo {author} {\bibfnamefont {F.~S.}\ \bibnamefont
  {{Guzm{\'a}n}}}, \ and\ \bibinfo {author} {\bibfnamefont {L.~A.}\
  \bibnamefont {{Ure{\~n}a-L{\'o}pez}}},\ }\href {\doibase
  10.1088/0264-9381/17/7/309} {\bibfield  {journal} {\bibinfo  {journal}
  {Classical and Quantum Gravity}\ }\textbf {\bibinfo {volume} {17}},\ \bibinfo
  {pages} {1707} (\bibinfo {year} {2000})},\ \Eprint
  {http://arxiv.org/abs/astro-ph/9908152} {arXiv:astro-ph/9908152 [astro-ph]}
  \BibitemShut {NoStop}%
\bibitem [{\citenamefont {{Guzm{\'a}n}}\ and\ \citenamefont
  {{Matos}}(2000)}]{Guzman2000}%
  \BibitemOpen
  \bibfield  {author} {\bibinfo {author} {\bibfnamefont {F.~S.}\ \bibnamefont
  {{Guzm{\'a}n}}}\ and\ \bibinfo {author} {\bibfnamefont {T.}~\bibnamefont
  {{Matos}}},\ }\href {\doibase 10.1088/0264-9381/17/1/102} {\bibfield
  {journal} {\bibinfo  {journal} {Classical and Quantum Gravity}\ }\textbf
  {\bibinfo {volume} {17}},\ \bibinfo {pages} {L9} (\bibinfo {year} {2000})},\
  \Eprint {http://arxiv.org/abs/gr-qc/9810028} {arXiv:gr-qc/9810028 [gr-qc]}
  \BibitemShut {NoStop}%
\bibitem [{\citenamefont {Goodman}(2000)}]{Goodman_2000}%
  \BibitemOpen
  \bibfield  {author} {\bibinfo {author} {\bibfnamefont {J.}~\bibnamefont
  {Goodman}},\ }\href {\doibase 10.1016/s1384-1076(00)00015-4} {\bibfield
  {journal} {\bibinfo  {journal} {New Astronomy}\ }\textbf {\bibinfo {volume}
  {5}},\ \bibinfo {pages} {103} (\bibinfo {year} {2000})}\BibitemShut {NoStop}%
\bibitem [{\citenamefont {Peebles}(2000)}]{Peebles_2000}%
  \BibitemOpen
  \bibfield  {author} {\bibinfo {author} {\bibfnamefont {P.~J.~E.}\
  \bibnamefont {Peebles}},\ }\href {\doibase 10.1086/312677} {\bibfield
  {journal} {\bibinfo  {journal} {The Astrophysical Journal}\ }\textbf
  {\bibinfo {volume} {534}},\ \bibinfo {pages} {L127} (\bibinfo {year}
  {2000})}\BibitemShut {NoStop}%
\bibitem [{\citenamefont {Amendola}\ and\ \citenamefont
  {Barbieri}(2006)}]{Amendola_2006}%
  \BibitemOpen
  \bibfield  {author} {\bibinfo {author} {\bibfnamefont {L.}~\bibnamefont
  {Amendola}}\ and\ \bibinfo {author} {\bibfnamefont {R.}~\bibnamefont
  {Barbieri}},\ }\href {\doibase 10.1016/j.physletb.2006.08.069} {\bibfield
  {journal} {\bibinfo  {journal} {Physics Letters B}\ }\textbf {\bibinfo
  {volume} {642}},\ \bibinfo {pages} {192} (\bibinfo {year}
  {2006})}\BibitemShut {NoStop}%
\bibitem [{\citenamefont {{Hwang}}\ and\ \citenamefont
  {{Noh}}(2009)}]{Hwang2009}%
  \BibitemOpen
  \bibfield  {author} {\bibinfo {author} {\bibfnamefont {J.-C.}\ \bibnamefont
  {{Hwang}}}\ and\ \bibinfo {author} {\bibfnamefont {H.}~\bibnamefont
  {{Noh}}},\ }\href {\doibase 10.1016/j.physletb.2009.08.031} {\bibfield
  {journal} {\bibinfo  {journal} {Physics Letters B}\ }\textbf {\bibinfo
  {volume} {680}},\ \bibinfo {pages} {1} (\bibinfo {year} {2009})},\ \Eprint
  {http://arxiv.org/abs/0902.4738} {arXiv:0902.4738 [astro-ph.CO]} \BibitemShut
  {NoStop}%
\bibitem [{\citenamefont {Marsh}(2016)}]{Marsh2016a}%
  \BibitemOpen
  \bibfield  {author} {\bibinfo {author} {\bibfnamefont {D.~J.~E.}\
  \bibnamefont {Marsh}},\ }\href {\doibase 10.1016/j.physrep.2016.06.005}
  {\bibfield  {journal} {\bibinfo  {journal} {Physics Reports}\ }\textbf
  {\bibinfo {volume} {643}},\ \bibinfo {pages} {1} (\bibinfo {year}
  {2016})}\BibitemShut {NoStop}%
\bibitem [{\citenamefont {Niemeyer}(2020)}]{NIEMEYER2020}%
  \BibitemOpen
  \bibfield  {author} {\bibinfo {author} {\bibfnamefont {J.~C.}\ \bibnamefont
  {Niemeyer}},\ }\href {\doibase https://doi.org/10.1016/j.ppnp.2020.103787}
  {\bibfield  {journal} {\bibinfo  {journal} {Progress in Particle and Nuclear
  Physics}\ }\textbf {\bibinfo {volume} {113}},\ \bibinfo {pages} {103787}
  (\bibinfo {year} {2020})}\BibitemShut {NoStop}%
\bibitem [{\citenamefont {Hui}\ \emph {et~al.}(2017)\citenamefont {Hui},
  \citenamefont {Ostriker}, \citenamefont {Tremaine},\ and\ \citenamefont
  {Witten}}]{Hui2017}%
  \BibitemOpen
  \bibfield  {author} {\bibinfo {author} {\bibfnamefont {L.}~\bibnamefont
  {Hui}}, \bibinfo {author} {\bibfnamefont {J.~P.}\ \bibnamefont {Ostriker}},
  \bibinfo {author} {\bibfnamefont {S.}~\bibnamefont {Tremaine}}, \ and\
  \bibinfo {author} {\bibfnamefont {E.}~\bibnamefont {Witten}},\ }\href
  {\doibase 10.1103/PhysRevD.95.043541} {\bibfield  {journal} {\bibinfo
  {journal} {Phys. Rev. D}\ }\textbf {\bibinfo {volume} {95}} (\bibinfo {year}
  {2017}),\ 10.1103/PhysRevD.95.043541},\ \Eprint
  {http://arxiv.org/abs/1610.08297} {arXiv:1610.08297} \BibitemShut {NoStop}%
\bibitem [{\citenamefont {Musoke}\ \emph {et~al.}(2020)\citenamefont {Musoke},
  \citenamefont {Hotchkiss},\ and\ \citenamefont {Easther}}]{Musoke:2019ima}%
  \BibitemOpen
  \bibfield  {author} {\bibinfo {author} {\bibfnamefont {N.}~\bibnamefont
  {Musoke}}, \bibinfo {author} {\bibfnamefont {S.}~\bibnamefont {Hotchkiss}}, \
  and\ \bibinfo {author} {\bibfnamefont {R.}~\bibnamefont {Easther}},\ }\href
  {\doibase 10.1103/PhysRevLett.124.061301} {\bibfield  {journal} {\bibinfo
  {journal} {Phys. Rev. Lett.}\ }\textbf {\bibinfo {volume} {124}},\ \bibinfo
  {pages} {061301} (\bibinfo {year} {2020})},\ \Eprint
  {http://arxiv.org/abs/1909.11678} {arXiv:1909.11678 [astro-ph.CO]}
  \BibitemShut {NoStop}%
\bibitem [{\citenamefont {Eggemeier}\ \emph {et~al.}(2021)\citenamefont
  {Eggemeier}, \citenamefont {Niemeyer},\ and\ \citenamefont
  {Easther}}]{Eggemeier:2020zeg}%
  \BibitemOpen
  \bibfield  {author} {\bibinfo {author} {\bibfnamefont {B.}~\bibnamefont
  {Eggemeier}}, \bibinfo {author} {\bibfnamefont {J.~C.}\ \bibnamefont
  {Niemeyer}}, \ and\ \bibinfo {author} {\bibfnamefont {R.}~\bibnamefont
  {Easther}},\ }\href {\doibase 10.1103/PhysRevD.103.063525} {\bibfield
  {journal} {\bibinfo  {journal} {Phys. Rev. D}\ }\textbf {\bibinfo {volume}
  {103}},\ \bibinfo {pages} {063525} (\bibinfo {year} {2021})},\ \Eprint
  {http://arxiv.org/abs/2011.13333} {arXiv:2011.13333 [astro-ph.CO]}
  \BibitemShut {NoStop}%
\bibitem [{\citenamefont {Guzm\'an}\ and\ \citenamefont {Ure\~na
  L\'opez}(2004)}]{Guzman2004}%
  \BibitemOpen
  \bibfield  {author} {\bibinfo {author} {\bibfnamefont {F.~S.}\ \bibnamefont
  {Guzm\'an}}\ and\ \bibinfo {author} {\bibfnamefont {L.~A.}\ \bibnamefont
  {Ure\~na L\'opez}},\ }\href {\doibase 10.1103/PhysRevD.69.124033} {\bibfield
  {journal} {\bibinfo  {journal} {Phys. Rev. D}\ }\textbf {\bibinfo {volume}
  {69}},\ \bibinfo {pages} {124033} (\bibinfo {year} {2004})}\BibitemShut
  {NoStop}%
\bibitem [{\citenamefont {Schwabe}\ \emph {et~al.}(2016)\citenamefont
  {Schwabe}, \citenamefont {Niemeyer},\ and\ \citenamefont
  {Engels}}]{Schwabe2016}%
  \BibitemOpen
  \bibfield  {author} {\bibinfo {author} {\bibfnamefont {B.}~\bibnamefont
  {Schwabe}}, \bibinfo {author} {\bibfnamefont {J.~C.}\ \bibnamefont
  {Niemeyer}}, \ and\ \bibinfo {author} {\bibfnamefont {J.~F.}\ \bibnamefont
  {Engels}},\ }\href {\doibase 10.1103/physrevd.94.043513} {\bibfield
  {journal} {\bibinfo  {journal} {Phys. Rev. D}\ }\textbf {\bibinfo {volume}
  {94}},\ \bibinfo {pages} {043513} (\bibinfo {year} {2016})}\BibitemShut
  {NoStop}%
\bibitem [{\citenamefont {Mocz}\ \emph {et~al.}(2017)\citenamefont {Mocz},
  \citenamefont {Vogelsberger}, \citenamefont {Robles}, \citenamefont {Zavala},
  \citenamefont {Boylan-Kolchin}, \citenamefont {Fialkov},\ and\ \citenamefont
  {Hernquist}}]{Mocz2017}%
  \BibitemOpen
  \bibfield  {author} {\bibinfo {author} {\bibfnamefont {P.}~\bibnamefont
  {Mocz}}, \bibinfo {author} {\bibfnamefont {M.}~\bibnamefont {Vogelsberger}},
  \bibinfo {author} {\bibfnamefont {V.~H.}\ \bibnamefont {Robles}}, \bibinfo
  {author} {\bibfnamefont {J.}~\bibnamefont {Zavala}}, \bibinfo {author}
  {\bibfnamefont {M.}~\bibnamefont {Boylan-Kolchin}}, \bibinfo {author}
  {\bibfnamefont {A.}~\bibnamefont {Fialkov}}, \ and\ \bibinfo {author}
  {\bibfnamefont {L.}~\bibnamefont {Hernquist}},\ }\href {\doibase
  10.1093/mnras/stx1887} {\bibfield  {journal} {\bibinfo  {journal} {Monthly
  Notices of the Royal Astronomical Society}\ }\textbf {\bibinfo {volume}
  {471}},\ \bibinfo {pages} {4559} (\bibinfo {year} {2017})}\BibitemShut
  {NoStop}%
\bibitem [{\citenamefont {Eggemeier}\ and\ \citenamefont
  {Niemeyer}(2019)}]{Eggemeier2019}%
  \BibitemOpen
  \bibfield  {author} {\bibinfo {author} {\bibfnamefont {B.}~\bibnamefont
  {Eggemeier}}\ and\ \bibinfo {author} {\bibfnamefont {J.~C.}\ \bibnamefont
  {Niemeyer}},\ }\href {\doibase 10.1103/PhysRevD.100.063528} {\bibfield
  {journal} {\bibinfo  {journal} {Phys. Rev. D}\ }\textbf {\bibinfo {volume}
  {100}},\ \bibinfo {pages} {063528} (\bibinfo {year} {2019})}\BibitemShut
  {NoStop}%
\bibitem [{\citenamefont {Lancaster}\ \emph {et~al.}(2020)\citenamefont
  {Lancaster}, \citenamefont {Giovanetti}, \citenamefont {Mocz}, \citenamefont
  {Kahn}, \citenamefont {Lisanti},\ and\ \citenamefont
  {Spergel}}]{Lancaster2019}%
  \BibitemOpen
  \bibfield  {author} {\bibinfo {author} {\bibfnamefont {L.}~\bibnamefont
  {Lancaster}}, \bibinfo {author} {\bibfnamefont {C.}~\bibnamefont
  {Giovanetti}}, \bibinfo {author} {\bibfnamefont {P.}~\bibnamefont {Mocz}},
  \bibinfo {author} {\bibfnamefont {Y.}~\bibnamefont {Kahn}}, \bibinfo {author}
  {\bibfnamefont {M.}~\bibnamefont {Lisanti}}, \ and\ \bibinfo {author}
  {\bibfnamefont {D.~N.}\ \bibnamefont {Spergel}},\ }\href {\doibase
  10.1088/1475-7516/2020/01/001} {\bibfield  {journal} {\bibinfo  {journal}
  {JCAP}\ }\textbf {\bibinfo {volume} {01}},\ \bibinfo {pages} {001} (\bibinfo
  {year} {2020})},\ \Eprint {http://arxiv.org/abs/1909.06381} {arXiv:1909.06381
  [astro-ph.CO]} \BibitemShut {NoStop}%
\bibitem [{\citenamefont {Landau}\ and\ \citenamefont
  {Lifshitz}(1977)}]{LandauCoulomb}%
  \BibitemOpen
  \bibfield  {author} {\bibinfo {author} {\bibfnamefont {L.~D.}\ \bibnamefont
  {Landau}}\ and\ \bibinfo {author} {\bibfnamefont {E.~M.}\ \bibnamefont
  {Lifshitz}},\ }\href@noop {} {\emph {\bibinfo {title} {Course of theoretical
  physics III: Quantum mechanics, Non-relativistic theory (3rd ed.)}}}\
  (\bibinfo {year} {1977})\BibitemShut {NoStop}%
\bibitem [{\citenamefont {Bar-Or}\ \emph {et~al.}(2018)\citenamefont {Bar-Or},
  \citenamefont {Fouvry},\ and\ \citenamefont {Tremaine}}]{Bar-Or2018}%
  \BibitemOpen
  \bibfield  {author} {\bibinfo {author} {\bibfnamefont {B.}~\bibnamefont
  {Bar-Or}}, \bibinfo {author} {\bibfnamefont {J.~B.}\ \bibnamefont {Fouvry}},
  \ and\ \bibinfo {author} {\bibfnamefont {S.}~\bibnamefont {Tremaine}},\
  }\href {\doibase 10.3847/1538-4357/aaf28c} {\bibfield  {journal} {\bibinfo
  {journal} {arXiv}\ } (\bibinfo {year} {2018}),\ 10.3847/1538-4357/aaf28c},\
  \Eprint {http://arxiv.org/abs/1809.07673} {arXiv:1809.07673} \BibitemShut
  {NoStop}%
\bibitem [{\citenamefont {Edwards}\ \emph
  {et~al.}(2018{\natexlab{a}})\citenamefont {Edwards}, \citenamefont {Kendall},
  \citenamefont {Hotchkiss},\ and\ \citenamefont {Easther}}]{Edwards2018}%
  \BibitemOpen
  \bibfield  {author} {\bibinfo {author} {\bibfnamefont {F.}~\bibnamefont
  {Edwards}}, \bibinfo {author} {\bibfnamefont {E.}~\bibnamefont {Kendall}},
  \bibinfo {author} {\bibfnamefont {S.}~\bibnamefont {Hotchkiss}}, \ and\
  \bibinfo {author} {\bibfnamefont {R.}~\bibnamefont {Easther}},\ }\href
  {\doibase 10.1088/1475-7516/2018/10/027} {\bibfield  {journal} {\bibinfo
  {journal} {J. Cosmol. Astropart. Phys.}\ }\textbf {\bibinfo {volume} {2018}}
  (\bibinfo {year} {2018}{\natexlab{a}}),\ 10.1088/1475-7516/2018/10/027},\
  \Eprint {http://arxiv.org/abs/1807.04037} {arXiv:1807.04037} \BibitemShut
  {NoStop}%
\bibitem [{\citenamefont {Marsh}\ and\ \citenamefont
  {Hoof}(2021)}]{Marsh:2021lqg}%
  \BibitemOpen
  \bibfield  {author} {\bibinfo {author} {\bibfnamefont {D.~J.~E.}\
  \bibnamefont {Marsh}}\ and\ \bibinfo {author} {\bibfnamefont
  {S.}~\bibnamefont {Hoof}},\ }\href@noop {} {\  (\bibinfo {year} {2021})},\
  \Eprint {http://arxiv.org/abs/2106.08797} {arXiv:2106.08797 [hep-ph]}
  \BibitemShut {NoStop}%
\bibitem [{\citenamefont {Rogers}\ and\ \citenamefont
  {Peiris}(2021)}]{Rogers:2020ltq}%
  \BibitemOpen
  \bibfield  {author} {\bibinfo {author} {\bibfnamefont {K.~K.}\ \bibnamefont
  {Rogers}}\ and\ \bibinfo {author} {\bibfnamefont {H.~V.}\ \bibnamefont
  {Peiris}},\ }\href {\doibase 10.1103/PhysRevLett.126.071302} {\bibfield
  {journal} {\bibinfo  {journal} {Phys. Rev. Lett.}\ }\textbf {\bibinfo
  {volume} {126}},\ \bibinfo {pages} {071302} (\bibinfo {year} {2021})},\
  \Eprint {http://arxiv.org/abs/2007.12705} {arXiv:2007.12705 [astro-ph.CO]}
  \BibitemShut {NoStop}%
\bibitem [{\citenamefont {Marsh}\ and\ \citenamefont
  {Niemeyer}(2019)}]{Marsh:2018zyw}%
  \BibitemOpen
  \bibfield  {author} {\bibinfo {author} {\bibfnamefont {D.~J.~E.}\
  \bibnamefont {Marsh}}\ and\ \bibinfo {author} {\bibfnamefont {J.~C.}\
  \bibnamefont {Niemeyer}},\ }\href {\doibase 10.1103/PhysRevLett.123.051103}
  {\bibfield  {journal} {\bibinfo  {journal} {Phys. Rev. Lett.}\ }\textbf
  {\bibinfo {volume} {123}},\ \bibinfo {pages} {051103} (\bibinfo {year}
  {2019})},\ \Eprint {http://arxiv.org/abs/1810.08543} {arXiv:1810.08543
  [astro-ph.CO]} \BibitemShut {NoStop}%
\bibitem [{\citenamefont {Kendall}\ and\ \citenamefont
  {Easther}(2020{\natexlab{a}})}]{Kendall:2019fep}%
  \BibitemOpen
  \bibfield  {author} {\bibinfo {author} {\bibfnamefont {E.}~\bibnamefont
  {Kendall}}\ and\ \bibinfo {author} {\bibfnamefont {R.}~\bibnamefont
  {Easther}},\ }\href {\doibase 10.1017/pasa.2020.3} {\bibfield  {journal}
  {\bibinfo  {journal} {Publ. Astron. Soc. Austral.}\ }\textbf {\bibinfo
  {volume} {37}},\ \bibinfo {pages} {e009} (\bibinfo {year}
  {2020}{\natexlab{a}})},\ \Eprint {http://arxiv.org/abs/1908.02508}
  {arXiv:1908.02508 [astro-ph.CO]} \BibitemShut {NoStop}%
\bibitem [{\citenamefont {Bar}\ \emph {et~al.}(2021)\citenamefont {Bar},
  \citenamefont {Blum},\ and\ \citenamefont {Sun}}]{bar2021}%
  \BibitemOpen
  \bibfield  {author} {\bibinfo {author} {\bibfnamefont {N.}~\bibnamefont
  {Bar}}, \bibinfo {author} {\bibfnamefont {K.}~\bibnamefont {Blum}}, \ and\
  \bibinfo {author} {\bibfnamefont {C.}~\bibnamefont {Sun}},\ }\href@noop {} {\
   (\bibinfo {year} {2021})},\ \Eprint {http://arxiv.org/abs/2111.03070}
  {arXiv:2111.03070 [hep-ph]} \BibitemShut {NoStop}%
\bibitem [{\citenamefont {Stott}\ and\ \citenamefont
  {Marsh}(2018)}]{Stott:2018opm}%
  \BibitemOpen
  \bibfield  {author} {\bibinfo {author} {\bibfnamefont {M.~J.}\ \bibnamefont
  {Stott}}\ and\ \bibinfo {author} {\bibfnamefont {D.~J.~E.}\ \bibnamefont
  {Marsh}},\ }\href {\doibase 10.1103/PhysRevD.98.083006} {\bibfield  {journal}
  {\bibinfo  {journal} {Phys. Rev. D}\ }\textbf {\bibinfo {volume} {98}},\
  \bibinfo {pages} {083006} (\bibinfo {year} {2018})},\ \Eprint
  {http://arxiv.org/abs/1805.02016} {arXiv:1805.02016 [hep-ph]} \BibitemShut
  {NoStop}%
\bibitem [{\citenamefont {Hlozek}\ \emph {et~al.}(2015)\citenamefont {Hlozek},
  \citenamefont {Grin}, \citenamefont {Marsh},\ and\ \citenamefont
  {Ferreira}}]{Hlozek:2014lca}%
  \BibitemOpen
  \bibfield  {author} {\bibinfo {author} {\bibfnamefont {R.}~\bibnamefont
  {Hlozek}}, \bibinfo {author} {\bibfnamefont {D.}~\bibnamefont {Grin}},
  \bibinfo {author} {\bibfnamefont {D.~J.~E.}\ \bibnamefont {Marsh}}, \ and\
  \bibinfo {author} {\bibfnamefont {P.~G.}\ \bibnamefont {Ferreira}},\ }\href
  {\doibase 10.1103/PhysRevD.91.103512} {\bibfield  {journal} {\bibinfo
  {journal} {Phys. Rev. D}\ }\textbf {\bibinfo {volume} {91}},\ \bibinfo
  {pages} {103512} (\bibinfo {year} {2015})},\ \Eprint
  {http://arxiv.org/abs/1410.2896} {arXiv:1410.2896 [astro-ph.CO]} \BibitemShut
  {NoStop}%
\bibitem [{\citenamefont {Dentler}\ \emph {et~al.}(2021)\citenamefont
  {Dentler}, \citenamefont {Marsh}, \citenamefont {Hlo\v{z}ek}, \citenamefont
  {Lagu\"e}, \citenamefont {Rogers},\ and\ \citenamefont
  {Grin}}]{Dentler:2021zij}%
  \BibitemOpen
  \bibfield  {author} {\bibinfo {author} {\bibfnamefont {M.}~\bibnamefont
  {Dentler}}, \bibinfo {author} {\bibfnamefont {D.~J.~E.}\ \bibnamefont
  {Marsh}}, \bibinfo {author} {\bibfnamefont {R.}~\bibnamefont {Hlo\v{z}ek}},
  \bibinfo {author} {\bibfnamefont {A.}~\bibnamefont {Lagu\"e}}, \bibinfo
  {author} {\bibfnamefont {K.~K.}\ \bibnamefont {Rogers}}, \ and\ \bibinfo
  {author} {\bibfnamefont {D.}~\bibnamefont {Grin}},\ }\href@noop {} {\
  (\bibinfo {year} {2021})},\ \Eprint {http://arxiv.org/abs/2111.01199}
  {arXiv:2111.01199 [astro-ph.CO]} \BibitemShut {NoStop}%
\bibitem [{\citenamefont {Ferreira}(2020)}]{Ferreira2020}%
  \BibitemOpen
  \bibfield  {author} {\bibinfo {author} {\bibfnamefont {E.~G.~M.}\
  \bibnamefont {Ferreira}},\ }\href {http://arxiv.org/abs/2005.03254} {\
  (\bibinfo {year} {2020})},\ \Eprint {http://arxiv.org/abs/2005.03254}
  {arXiv:2005.03254} \BibitemShut {NoStop}%
\bibitem [{\citenamefont {{Schive}}\ \emph {et~al.}(2014)\citenamefont
  {{Schive}}, \citenamefont {{Chiueh}},\ and\ \citenamefont
  {{Broadhurst}}}]{Schive2014}%
  \BibitemOpen
  \bibfield  {author} {\bibinfo {author} {\bibfnamefont {H.-Y.}\ \bibnamefont
  {{Schive}}}, \bibinfo {author} {\bibfnamefont {T.}~\bibnamefont {{Chiueh}}},
  \ and\ \bibinfo {author} {\bibfnamefont {T.}~\bibnamefont {{Broadhurst}}},\
  }\href {\doibase 10.1038/nphys2996} {\bibfield  {journal} {\bibinfo
  {journal} {Nature Physics}\ }\textbf {\bibinfo {volume} {10}},\ \bibinfo
  {pages} {496} (\bibinfo {year} {2014})},\ \Eprint
  {http://arxiv.org/abs/1406.6586} {arXiv:1406.6586} \BibitemShut {NoStop}%
\bibitem [{\citenamefont {Kendall}\ and\ \citenamefont
  {Easther}(2020{\natexlab{b}})}]{Kendall2020a}%
  \BibitemOpen
  \bibfield  {author} {\bibinfo {author} {\bibfnamefont {E.}~\bibnamefont
  {Kendall}}\ and\ \bibinfo {author} {\bibfnamefont {R.}~\bibnamefont
  {Easther}},\ }\href {\doibase 10.1017/pasa.2020.3} {\bibfield  {journal}
  {\bibinfo  {journal} {Publ. Astron. Soc. Aust.}\ } (\bibinfo {year}
  {2020}{\natexlab{b}}),\ 10.1017/pasa.2020.3},\ \Eprint
  {http://arxiv.org/abs/1908.02508} {arXiv:1908.02508} \BibitemShut {NoStop}%
\bibitem [{\citenamefont {{Chandrasekhar}}(1943)}]{Chandrasekhar}%
  \BibitemOpen
  \bibfield  {author} {\bibinfo {author} {\bibfnamefont {S.}~\bibnamefont
  {{Chandrasekhar}}},\ }\href {\doibase 10.1086/144517} {\bibfield  {journal}
  {\bibinfo  {journal} {\apj}\ }\textbf {\bibinfo {volume} {97}},\ \bibinfo
  {pages} {255} (\bibinfo {year} {1943})}\BibitemShut {NoStop}%
\bibitem [{\citenamefont {Edwards}\ \emph
  {et~al.}(2018{\natexlab{b}})\citenamefont {Edwards}, \citenamefont {Kendall},
  \citenamefont {Hotchkiss},\ and\ \citenamefont {Easther}}]{Edwards:2018ccc}%
  \BibitemOpen
  \bibfield  {author} {\bibinfo {author} {\bibfnamefont {F.}~\bibnamefont
  {Edwards}}, \bibinfo {author} {\bibfnamefont {E.}~\bibnamefont {Kendall}},
  \bibinfo {author} {\bibfnamefont {S.}~\bibnamefont {Hotchkiss}}, \ and\
  \bibinfo {author} {\bibfnamefont {R.}~\bibnamefont {Easther}},\ }\href
  {\doibase 10.1088/1475-7516/2018/10/027} {\bibfield  {journal} {\bibinfo
  {journal} {JCAP}\ }\textbf {\bibinfo {volume} {10}},\ \bibinfo {pages} {027}
  (\bibinfo {year} {2018}{\natexlab{b}})},\ \Eprint
  {http://arxiv.org/abs/1807.04037} {arXiv:1807.04037 [astro-ph.CO]}
  \BibitemShut {NoStop}%
\bibitem [{\citenamefont {Plummer}(1911)}]{Plummer1911}%
  \BibitemOpen
  \bibfield  {author} {\bibinfo {author} {\bibfnamefont {H.~C.}\ \bibnamefont
  {Plummer}},\ }\href {\doibase 10.1093/mnras/71.5.460} {\bibfield  {journal}
  {\bibinfo  {journal} {Mon. Not. R. Astron. Soc.}\ }\textbf {\bibinfo {volume}
  {71}},\ \bibinfo {pages} {460} (\bibinfo {year} {1911})}\BibitemShut
  {NoStop}%
\bibitem [{\citenamefont {Schive}\ \emph {et~al.}(2014)\citenamefont {Schive},
  \citenamefont {Liao}, \citenamefont {Woo}, \citenamefont {Wong},
  \citenamefont {Chiueh}, \citenamefont {Broadhurst},\ and\ \citenamefont
  {Hwang}}]{CoreHalo}%
  \BibitemOpen
  \bibfield  {author} {\bibinfo {author} {\bibfnamefont {H.-Y.}\ \bibnamefont
  {Schive}}, \bibinfo {author} {\bibfnamefont {M.-H.}\ \bibnamefont {Liao}},
  \bibinfo {author} {\bibfnamefont {T.-P.}\ \bibnamefont {Woo}}, \bibinfo
  {author} {\bibfnamefont {S.-K.}\ \bibnamefont {Wong}}, \bibinfo {author}
  {\bibfnamefont {T.}~\bibnamefont {Chiueh}}, \bibinfo {author} {\bibfnamefont
  {T.}~\bibnamefont {Broadhurst}}, \ and\ \bibinfo {author} {\bibfnamefont
  {W.-Y.~P.}\ \bibnamefont {Hwang}},\ }\href {\doibase
  10.1103/PhysRevLett.113.261302} {\bibfield  {journal} {\bibinfo  {journal}
  {Phys. Rev. Lett.}\ }\textbf {\bibinfo {volume} {113}},\ \bibinfo {pages}
  {261302} (\bibinfo {year} {2014})}\BibitemShut {NoStop}%
\bibitem [{\citenamefont {Helfer}\ \emph {et~al.}(2017)\citenamefont {Helfer},
  \citenamefont {Marsh}, \citenamefont {Clough}, \citenamefont {Fairbairn},
  \citenamefont {Lim},\ and\ \citenamefont {Becerril}}]{Helfer2017}%
  \BibitemOpen
  \bibfield  {author} {\bibinfo {author} {\bibfnamefont {T.}~\bibnamefont
  {Helfer}}, \bibinfo {author} {\bibfnamefont {D.~J.}\ \bibnamefont {Marsh}},
  \bibinfo {author} {\bibfnamefont {K.}~\bibnamefont {Clough}}, \bibinfo
  {author} {\bibfnamefont {M.}~\bibnamefont {Fairbairn}}, \bibinfo {author}
  {\bibfnamefont {E.~A.}\ \bibnamefont {Lim}}, \ and\ \bibinfo {author}
  {\bibfnamefont {R.}~\bibnamefont {Becerril}},\ }\href {\doibase
  10.1088/1475-7516/2017/03/055} {\bibfield  {journal} {\bibinfo  {journal}
  {Journal of Cosmology and Astroparticle Physics}\ }\textbf {\bibinfo {volume}
  {2017}},\ \bibinfo {pages} {055–055} (\bibinfo {year} {2017})}\BibitemShut
  {NoStop}%
\bibitem [{\citenamefont {Dutta~Chowdhury}\ \emph {et~al.}(2021)\citenamefont
  {Dutta~Chowdhury}, \citenamefont {van~den Bosch}, \citenamefont {Robles},
  \citenamefont {van Dokkum}, \citenamefont {Schive}, \citenamefont {Chiueh},\
  and\ \citenamefont {Broadhurst}}]{YaleDDC}%
  \BibitemOpen
  \bibfield  {author} {\bibinfo {author} {\bibfnamefont {D.}~\bibnamefont
  {Dutta~Chowdhury}}, \bibinfo {author} {\bibfnamefont {F.~C.}\ \bibnamefont
  {van~den Bosch}}, \bibinfo {author} {\bibfnamefont {V.~H.}\ \bibnamefont
  {Robles}}, \bibinfo {author} {\bibfnamefont {P.}~\bibnamefont {van Dokkum}},
  \bibinfo {author} {\bibfnamefont {H.-Y.}\ \bibnamefont {Schive}}, \bibinfo
  {author} {\bibfnamefont {T.}~\bibnamefont {Chiueh}}, \ and\ \bibinfo {author}
  {\bibfnamefont {T.}~\bibnamefont {Broadhurst}},\ }\href {\doibase
  10.3847/1538-4357/ac043f} {\bibfield  {journal} {\bibinfo  {journal} {The
  Astrophysical Journal}\ }\textbf {\bibinfo {volume} {916}},\ \bibinfo {pages}
  {27} (\bibinfo {year} {2021})}\BibitemShut {NoStop}%
\bibitem [{\citenamefont {Annulli}\ \emph {et~al.}(2020)\citenamefont
  {Annulli}, \citenamefont {Cardoso},\ and\ \citenamefont {Vicente}}]{LisbonA}%
  \BibitemOpen
  \bibfield  {author} {\bibinfo {author} {\bibfnamefont {L.}~\bibnamefont
  {Annulli}}, \bibinfo {author} {\bibfnamefont {V.}~\bibnamefont {Cardoso}}, \
  and\ \bibinfo {author} {\bibfnamefont {R.}~\bibnamefont {Vicente}},\ }\href
  {\doibase 10.1103/PhysRevD.102.063022} {\bibfield  {journal} {\bibinfo
  {journal} {Phys. Rev. D}\ }\textbf {\bibinfo {volume} {102}},\ \bibinfo
  {pages} {063022} (\bibinfo {year} {2020})},\ \Eprint
  {http://arxiv.org/abs/2009.00012} {arXiv:2009.00012 [gr-qc]} \BibitemShut
  {NoStop}%
\bibitem [{\citenamefont {Burke-Spolaor}\ \emph {et~al.}(2019)\citenamefont
  {Burke-Spolaor} \emph {et~al.}}]{Burke-Spolaor:2018bvk}%
  \BibitemOpen
  \bibfield  {author} {\bibinfo {author} {\bibfnamefont {S.}~\bibnamefont
  {Burke-Spolaor}} \emph {et~al.},\ }\href {\doibase 10.1007/s00159-019-0115-7}
  {\bibfield  {journal} {\bibinfo  {journal} {Astron. Astrophys. Rev.}\
  }\textbf {\bibinfo {volume} {27}},\ \bibinfo {pages} {5} (\bibinfo {year}
  {2019})},\ \Eprint {http://arxiv.org/abs/1811.08826} {arXiv:1811.08826
  [astro-ph.HE]} \BibitemShut {NoStop}%
\bibitem [{\citenamefont {Amaro-Seoane}\ \emph {et~al.}(2017)\citenamefont
  {Amaro-Seoane} \emph {et~al.}}]{LISA:2017pwj}%
  \BibitemOpen
  \bibfield  {author} {\bibinfo {author} {\bibfnamefont {P.}~\bibnamefont
  {Amaro-Seoane}} \emph {et~al.} (\bibinfo {collaboration} {LISA}),\
  }\href@noop {} {\  (\bibinfo {year} {2017})},\ \Eprint
  {http://arxiv.org/abs/1702.00786} {arXiv:1702.00786 [astro-ph.IM]}
  \BibitemShut {NoStop}%
\bibitem [{\citenamefont {Milosavljevic}\ and\ \citenamefont
  {Merritt}(2003)}]{Milosavljevic:2002bn}%
  \BibitemOpen
  \bibfield  {author} {\bibinfo {author} {\bibfnamefont {M.}~\bibnamefont
  {Milosavljevic}}\ and\ \bibinfo {author} {\bibfnamefont {D.}~\bibnamefont
  {Merritt}},\ }\href {\doibase 10.1086/378086} {\bibfield  {journal} {\bibinfo
   {journal} {Astrophys. J.}\ }\textbf {\bibinfo {volume} {596}},\ \bibinfo
  {pages} {860} (\bibinfo {year} {2003})},\ \Eprint
  {http://arxiv.org/abs/astro-ph/0212459} {arXiv:astro-ph/0212459} \BibitemShut
  {NoStop}%
\bibitem [{\citenamefont {Barausse}\ and\ \citenamefont
  {Lapi}(2020)}]{Barausse:2020gbp}%
  \BibitemOpen
  \bibfield  {author} {\bibinfo {author} {\bibfnamefont {E.}~\bibnamefont
  {Barausse}}\ and\ \bibinfo {author} {\bibfnamefont {A.}~\bibnamefont
  {Lapi}},\ }\href@noop {} {\  (\bibinfo {year} {2020})},\ \Eprint
  {http://arxiv.org/abs/2011.01994} {arXiv:2011.01994 [astro-ph.GA]}
  \BibitemShut {NoStop}%
\bibitem [{\citenamefont {Zagorac}\ \emph {et~al.}(2021)\citenamefont
  {Zagorac}, \citenamefont {Sands}, \citenamefont {Padmanabhan},\ and\
  \citenamefont {Easther}}]{Zagorac:2021qxq}%
  \BibitemOpen
  \bibfield  {author} {\bibinfo {author} {\bibfnamefont {J.~L.}\ \bibnamefont
  {Zagorac}}, \bibinfo {author} {\bibfnamefont {I.}~\bibnamefont {Sands}},
  \bibinfo {author} {\bibfnamefont {N.}~\bibnamefont {Padmanabhan}}, \ and\
  \bibinfo {author} {\bibfnamefont {R.}~\bibnamefont {Easther}},\ }\href@noop
  {} {\  (\bibinfo {year} {2021})},\ \Eprint {http://arxiv.org/abs/2109.01920}
  {arXiv:2109.01920 [astro-ph.CO]} \BibitemShut {NoStop}%
\bibitem [{\citenamefont {Schwabe}\ \emph {et~al.}(2020)\citenamefont
  {Schwabe}, \citenamefont {Gosenca}, \citenamefont {Behrens}, \citenamefont
  {Niemeyer},\ and\ \citenamefont {Easther}}]{Schwabe2020}%
  \BibitemOpen
  \bibfield  {author} {\bibinfo {author} {\bibfnamefont {B.}~\bibnamefont
  {Schwabe}}, \bibinfo {author} {\bibfnamefont {M.}~\bibnamefont {Gosenca}},
  \bibinfo {author} {\bibfnamefont {C.}~\bibnamefont {Behrens}}, \bibinfo
  {author} {\bibfnamefont {J.~C.}\ \bibnamefont {Niemeyer}}, \ and\ \bibinfo
  {author} {\bibfnamefont {R.}~\bibnamefont {Easther}},\ }\href
  {http://arxiv.org/abs/2007.08256} {\bibfield  {journal} {\bibinfo  {journal}
  {Phys. Rev. D.}\ ,\ \bibinfo {pages} {1}} (\bibinfo {year} {2020})},\ \Eprint
  {http://arxiv.org/abs/2007.08256} {arXiv:2007.08256} \BibitemShut {NoStop}%
\end{thebibliography}%

\end{document}